\documentclass[prb, twocolumn, showpacs, amsmath, amssymb,superscriptaddress]{revtex4-1}

\usepackage{amssymb,amsfonts,amsmath} 
\usepackage{graphicx}
\usepackage{epsfig}
\usepackage{epstopdf}
\usepackage{color}
\usepackage{hyperref}
\usepackage{longtable}
\usepackage{bbm}
\usepackage{ulem}
\begin{document} 

\title{Non-equilibrium Optical Conductivity: General Theory and Application to Transient Phases}

\author{D.M.\ Kennes}
\affiliation{Department of Physics, Columbia University, New York, NY 10027, USA}
\author{E.Y.\ Wilner}
\affiliation{Department of Physics, Columbia University, New York, NY 10027, USA}
\author{D.R.\ Reichman}
\affiliation{Department of Chemistry, Columbia University, New York, NY 10027, USA}
\author{A.J.\ Millis}
\affiliation{Department of Physics, Columbia University, New York, NY 10027, USA}

\begin{abstract} 
A non-equilibrium theory of optical conductivity of  dirty-limit superconductors and commensurate charge density wave is presented. We discuss the current response to different experimentally relevant light-field probe pulses and show that a single frequency definition of the optical conductivity $\sigma(\omega)\equiv j(\omega)/E(\omega)$ is difficult to interpret out of the adiabatic limit. We identify characteristic time domain signatures distinguishing between superconducting, normal metal and charge density wave states. We also suggest a route to directly address the instantaneous superfluid stiffness of a superconductor by shaping the probe light field.   
\end{abstract}

\date{\today} 
\maketitle
\section{Introduction}
`Ultrafast' (typically optical or infrared) radiation pulses are now being employed to modify the electronic properties of materials including ferroelectrics, manganites, Mott insulators, and cuprate and organic superconductors.\cite{Garret97,Merlin97,Rini07,Wall09,Forst11,Fausti11,Matsunaga12,Matsunaga13,Matsunaga14,Hu14,Mankowsky15,Singla15,He16,Mitrano16,Nicoletti16} In these experiments the electronic state is typically studied via the response to an incident `probe' radiation field; the probe pulse is applied over a fixed and often relatively short time interval and the response is recorded in the time domain.  While a general formulation of nonequilibrium response exists as part of the Baym-Kadanoff-Keldysh nonequilibrium quantum field theory,\cite{Rammer07,Jauchobook} and the theory of the nonequilibrium properties of moderately perturbed conventional superconductors has been extensively studied,\cite{Rammer07} an extension of the theory to the case of transient phases such as charge density or superconducting order appears to be lacking. The calculation in the transient case is complicated by the fact that  time-translation invariance is broken, so that the nonequilibrium conductivity is a function of two frequencies which, as recently emphasized by Orenstein and Dodge, cannot be unambiguously collapsed to a function of a single frequency. \cite{Orenstein16,Nicoletticomment,pertfreeref1,pertfreeref2}    

In this paper we present a calculation of the nonequilibrium  optical response of a material with transient charge density or superconducting order (for a study of the optical properties of normal states see, e.g., Ref.~\onlinecite{condnorm1} and \onlinecite{condnorm2}). Our calculation uses a modified version of a time-dependent mean field approximation to treat the transient superconductivity or charge density order  and incorporates physically reasonable \cite{Quenchlit} pulse shapes for the buildup and decay of superconducting and density wave order. For simplicity our analysis is restricted to the `dirty limit' in which the basic electronic scattering rate is large compared to the gap.   Generalizing and extending the analysis of Refs.~\onlinecite{Orenstein16,Nicoletticomment}  we present time-domain signatures of the important physics, and show how appropriately tailored pulses can directly reveal the superfluid stiffness.  We demonstrate the pitfalls of collapsing the non-equilibrium conductivity to a single function of frequency. 


The rest of the paper is organized as follows. In section \ref{theory} we review  the conventional (Keldysh) theory of linear response in the time domain, in section \ref{examples} we present some simple equilibrium examples that reveal the essential features of the superconducting and charge density wave response and guide the interpretation of our calculations. Section \ref{details} gives the specifics of our calculation of the dirty limit conductivity and in sections \ref{resultstime} and \ref{resultsfreq} we present and discuss our non-equilibrium results. Section \ref{summary} is a summary and conclusion.  

\section{Linear Response \label{theory}}
We are interested in  a generic, not necessarily equilibrium, system, to which a weak `probe' electric field $E_{\rm probe}(t)$ is applied. The probe field will generate  an additional current $\delta j(t)$ (Note that out-of-equilibrium a system may have a current $j$ even if $E_{\rm probe}=0$; here we are interested only in the current attributable to the probe field, and we do not address the experimental issues involved in empirically defining and measuring $\delta j$). We suppose that the probe field is sufficiently weak that the response is defined by a causal linear response function $\sigma(t,t^\prime)$ as   
\begin{equation}
\delta j(t)=\int\limits_{-\infty}^\infty \sigma(t,t^\prime) E_{\rm probe}(t^\prime) dt^\prime.
\label{sigmadef}
\end{equation} 
Causality implies that $\sigma(t,t^\prime)=0$ if $t^\prime>t$.

In a non-superconducting material, currents typically decay with time so that $\lim_{t\rightarrow\infty}\sigma(t,t^\prime)=0$ whereas the dissipationless superfluid response of  a superconductor  implies that $\lim_{t\rightarrow\infty}\sigma(t,t^\prime)=\rho_S>0$ defining the superfluid stiffness $\rho_S$ so that for a superconductor we may write
\begin{equation}
\sigma(t,t^\prime)=\rho_S\Theta(t-t^\prime)+\sigma_{reg}(t,t^\prime)
\label{rhosdef}
\end{equation}
where $\sigma_{reg}(t,t^\prime)$ vanishes as $t\rightarrow\infty$.

It is often convenient to formulate calculations  in terms of the vector potential $A$. We are interested in the response to transverse fields for which $E=-\partial_tA$ (we use units in which $c=1$), Defining the current-current response $\chi_{JJ}(t,t^\prime)=-\partial_{t^\prime}\sigma(t,t^\prime)$, integrating by parts and assuming that $A(t^\prime\to-\infty)=0$ we obtain
\begin{equation}
\delta j(t)=\int\limits_{-\infty}^\infty \left[K(t)\delta(t-t^\prime)-\chi_{JJ}(t,t^\prime) \right]A_{\rm probe}(t^\prime) dt^\prime.
\label{chJJdef}
\end{equation} 
with ``kinetic energy'' 
\begin{equation}
K(t)=\sigma(t,t),
\label{Kdef1}
\end{equation}
where in Eq.~\eqref{Kdef1}  the equal time conductivity is defined as the limit in which $t^\prime$ approaches $t$ from below. Equation~\eqref{Kdef1} is the familiar conductance sum rule.


The calculation of $K$ and $\chi$ may be formulated in terms of the Keldysh two time contour Green's functions, 
\begin{align}
G^{\text{R}}(t,t')&=-i \Theta(t-t') \left\langle  \big [\Psi[t], \Psi^{\dagger}[t']\big ]_+\right \rangle,\\
G^{\text{K}}(t,t')&=-i  \left\langle  \big [\Psi[t], \Psi^{\dagger}[t']\big]_-\right \rangle,\\
G^{\text{A}}(t,t')&=i \Theta(t'-t) \left\langle \big  [\Psi[t], \Psi^{\dagger}[t']\big]_+\right \rangle,
\end{align}
where $\Psi^\dagger$ is an electron creation operator, space indices have been suppressed, the subscript $\pm$ denotes anticommutation or commutation respectively, the time dependence is computed with respect to the full Hamiltonian $\hat{H}$ and the expectation values $\left\langle\dots\right\rangle$ are taken with respect to an initial density matrix.  

We have in general
\begin{equation}
j(t)=\frac{1}{2}Tr\left[\hat{J}_{op}\left(1-iG^K(t,t)\right)\right],
\label{jdef}
\end{equation}
with $\hat{J}_{op}=\delta\hat{H}/\delta{A}$ the gauge invariant current operator.  Linearizing in $A_{\rm probe}$ we find Eq.~\eqref{chJJdef} with 
\begin{equation}
K(t)=\frac{1}{2}Tr\left[\frac{\delta^2\hat{H}}{\delta A^2}\left(1-iG^K(t,t)\right)\right]
\label{Kdef2}
\end{equation} and
\begin{equation}
\chi_{JJ}(t,t^\prime)=\frac{-i}{2}Tr\left[\hat{J}_{op}\frac{\delta G^K(t,t)}{\delta A(t^\prime)}\right],
\label{chidef}
\end{equation}
where $\frac{\delta G^K(t,t)}{\delta A(t^\prime)}$ is the functional derivative of  $G^K$ with respect to the probe vector potential.

In this paper we shall primarily be interested in situations (in particular variants of the time dependent mean field approximation) in which vertex corrections can be neglected, in which case 

\begin{eqnarray}
\frac{\delta G^K(t,t)}{\delta A(t^\prime)}&=&\int dt^\prime\bigg[G^R(t,t^\prime)\hat{J}_{op}(t^\prime)G^K(t^\prime,t)
\nonumber \\
&&+ G^K(t,t^\prime)\hat{J}_{op}(t^\prime)G^A(t^\prime,t)\bigg]A(t^\prime),
\end{eqnarray}
so 
\begin{eqnarray}
\chi_{JJ}(t,t^\prime)&=&\frac{-i}{2}Tr\bigg[\hat{J}_{op}G^R(t,t^\prime)\hat{J}_{op}G^K(t^\prime,t)
\nonumber \\
&&\hspace{0.3in}+ \hat{J}_{op}G^K(t,t^\prime)\hat{J}_{op}G^A(t^\prime,t)\bigg].
\label{chidef1}
\end{eqnarray}

\section{Simple Equilibrium examples \label{examples}}
In equilibrium $\sigma(t,t^\prime)$ and $\chi_{JJ}(t,t^\prime)$ are functions only of the time difference $\tau=t-t^\prime$. In the simple Drude model of a normal metal with scattering rate $\gamma$ and total spectral weight $K$  we have
\begin{equation}
\sigma_D(\tau)=Ke^{-\gamma \tau}\Theta(\tau).
\label{sigmaD}
\end{equation}
The defining time-domain feature of a normal metal is a decay of induced current on a time scale set by the scattering rate $\gamma$.

In an s-wave BCS superconductor with gap $\Delta$ and superfluid stiffness $\rho_S$ at temperature $T=0$ we have
\begin{equation}
\sigma_{SC}(\tau)=\rho_S\Theta(\tau)+\sigma_{\rm reg}(\tau)
\label{sigmasc}
\end{equation}
where $\sigma_{\rm reg}(\tau)$ vanishes as $\tau\rightarrow\infty$. The defining time-domain feature of a superconductor is that the current induced by an electric field pulse persists to $t\rightarrow\infty$. 

A case of particular interest is the `dirty'  (Mattis-Bardeen) limit, in which the superconducting gap is very small compared to the scattering rate $\gamma$. In this case it is convenient to write
\begin{equation}
\sigma_{reg}(\tau)=\sigma_D(\tau)+\sigma^{\rm MB}_{\rm reg}(\tau)
\label{sdef}
\end{equation}
because up to corrections of order $2\Delta/\gamma$, which we neglect in the dirty limit, $\sigma^{\rm MB}_{\rm reg}$ is the product of the superfluid stiffness $\rho_S=\pi K\Delta/(2\gamma)$ and a function $s$ that only depends on $2\Delta \tau$. $s(\tau=0)=-1$ and $s$ becomes very small for $\tau>1/\Delta$ so the total conductivity (equivalent to the current produced by a delta-function in time $E$-field pulse) is
\begin{equation}
\sigma(\tau)=\frac{K}{\gamma}\left\{\gamma e^{-\gamma \tau}+\frac{\pi \Delta}{2}\left[1+s(2\Delta \tau)\right]\right\}
\label{sigmasc1},
\end{equation}
with $s(t)=2/\pi\int_0^\infty dx\cos(x t) (s(x) - 1)$ and 
\begin{equation}
s(x)=\begin{cases}
 0&x\leq 1\\
 \left(1+\frac{1}{x}\right)E\left(\frac{x-1}{x+1}\right)-\frac{2}{x}K\left(\frac{x-1}{x+1}\right) & x >1
\end{cases}.\label{sigmasc2}
\end{equation}
$E$ and $K$ are complete elliptic integrals. 

In the above theory of a superconductor in the dirty limit the superfluid stiffness $\rho_S$ is proportional to the gap $\Delta$. To disentangle contributions of the two we also introduce a theory in which the superfluid stiffness and gap can be tuned independently. For this we define a fraction $0\leq A\leq 1$ which shifts weight from the superfluid (constant) part to a regular contribution at oscillatory frequency $2\Delta$
\begin{align}
\sigma(\tau)=&\frac{K}{\gamma}\bigg\{\gamma e^{-\gamma \tau}+\frac{\pi \Delta}{2}\bigg[A+s(2\Delta \tau)\notag\\
&+(1-A)\frac{\cos(2\Delta\tau) +2\Delta \tau\sin(2\Delta\tau))}{(2\Delta\tau)^2 + 1}\bigg]\bigg\}
\label{sigmasc3}.
\end{align}
For $A=1$ Eq.~\eqref{sigmasc3} reduces to Eq.~\eqref{sigmasc1}.

We contrast the superconducting state to the charge density wave state. The conductivity in the commensurate (nested Fermi surface) charge density wave state can be written in a similar form to Eq.~\eqref{sigmasc1} 
\begin{equation}
\sigma(\tau)=\frac{K}{\gamma}\left\{\gamma e^{-\gamma \tau}+\frac{\pi \Delta}{2}c(2\Delta \tau)\right\}
\label{sigmacdw1},
\end{equation}
but with the modified function 
\begin{equation}
c(x)=\begin{cases}
 0&x\leq 1\\
 \frac{\left(x^2+1\right)\left(\beta E(\alpha)+\left(\alpha^2-\beta\right)K(\alpha)\right)}{x(x+1)\alpha^2} & x >1
\end{cases},\label{sigmacdw2}
\end{equation} 
with
\begin{align}
\alpha&=\frac{x-1}{x+1}\\
\beta&=\frac{(x-1)^2}{x^2+1}
\end{align}
and the superfluid stiffness absent. 

\begin{figure}
\begin{center} 
\includegraphics[width=\columnwidth, angle=-0]{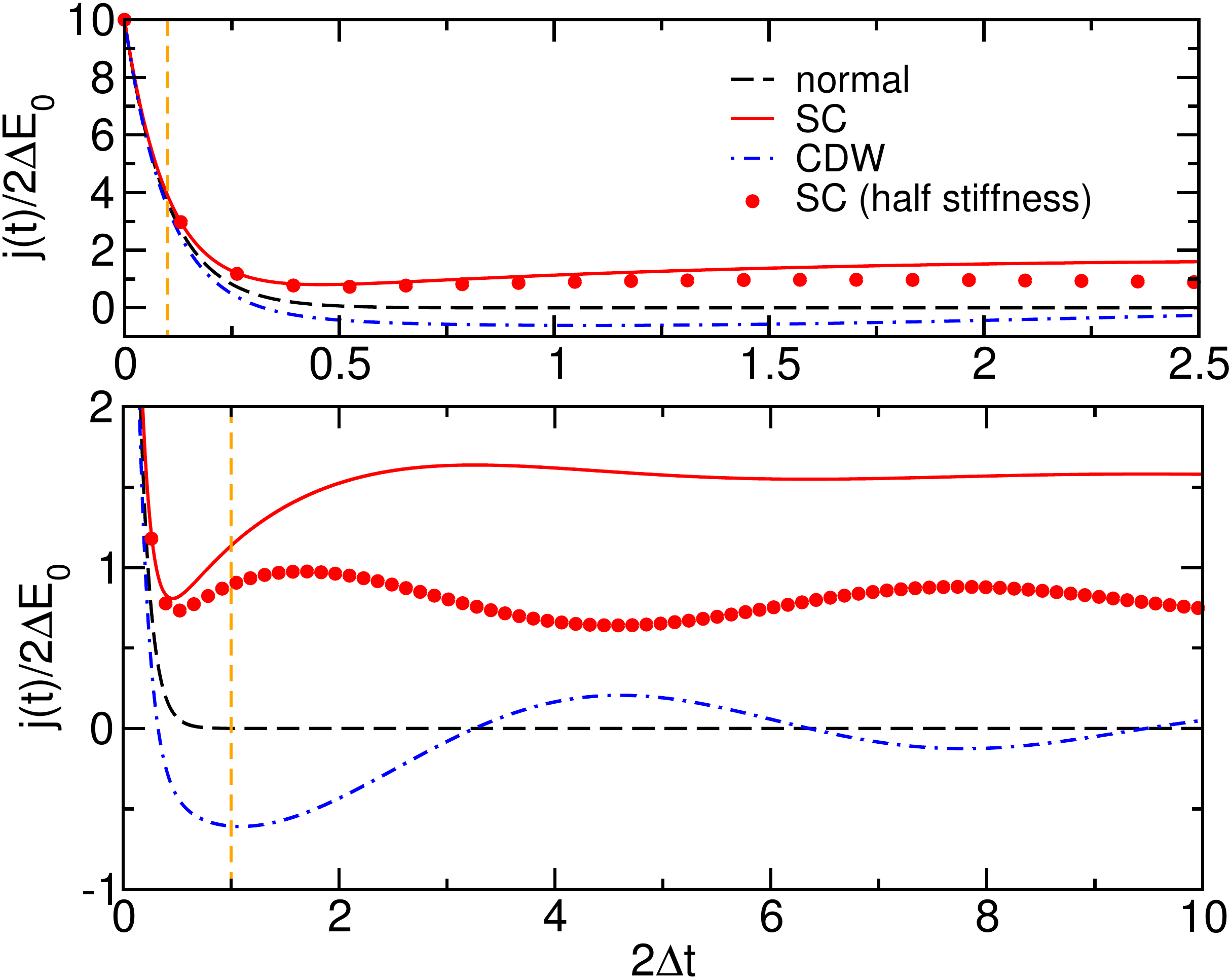}
\caption{Superconducting (solid line $A=1$ and dots $A=0.5$), charge density wave (dashed-dotted line) and normal (dashed line) currents produced by a delta function electric field pulse $E(t)=E_0\delta(t)$ calculated from Eqs. \eqref{sigmasc3}, \eqref{sigmacdw1} and \eqref{sigmaD}, respectively, plotted vs time in units of $(2\Delta)^{-1}$. The scattering rate is chosen as $\gamma/2\Delta=10$ and $K/\gamma=1$. The upper and lower panel show different time and current scales. The upper panel shows that the initial exponential decay is set by a timescale $\sim 1/\gamma$ (orange dashed vertical line). The lower panel indicates that the build up of the supercurrent in the superconducting state and the oscillations in the charge density as well as the superconducting state around their asymptotic value are governed by a time scale $\sim 1/\Delta$ (orange dashed vertical line).
}  
\label{samplecond}
\end{center}
\end{figure}

Fig.~\ref{samplecond} shows the current response $j(t)=\int_{-\infty}^tdt'\;\sigma(t-t')E(t')$  to a delta-probe field $E(t)=E_0\delta(t)$ (thus $A=-E_0\Theta(t)$). Note that this probe field cannot be applied in practice, but the response yields valuable intuition.  We show results for (i) the superconducting state current calculated from the conductivity Eqs.~\eqref{sigmasc3} and \eqref{sigmasc2} (red solid line BCS limit $A=1$ and dots reduced superfluid stiffness case, $A=0.5$), (ii) the normal state current  obtained by Eq.~\eqref{sigmaD} (dashed line) and (iii) the charge density wave state current evaluated from the conductivity  Eqs.~\eqref{sigmacdw1} and \eqref{sigmacdw2} with $\gamma/2\Delta=10$, where the basic unit of time is $(2\Delta)^{-1}$. The normal state response decays rapidly with time with a rate set by $\gamma$. The same initial fast decay can be found in the superconducting and charge density wave state as the additive drude contributions decays on a time scale $\sim 1/\gamma$. The superconducting response builds up over a time of the order of the inverse gap before saturating at the value prescribed by the superfluid stiffness. The evolution towards the finite supercurrent is superimposed by oscillations with frequency set by the gap. These oscillations are stronger in the $A<1$ case, as weight is shifted from the stiffness to the oscillatory contribution.  The current in the charge density wave state behaves opposite to the superconducting state, falling below the normal current at short times displaying oscillatory convergence to zero at long times. The oscillation frequency is again $\sim \Delta$.    

In many time-domain experiments,\cite{Averitt00,Larsen11,Matsunaga12}  the protocol is to apply an electric field pulse which integrates to zero of approximately the form  (we also give the corresponding vector potential)
\begin{align}
E_{\rm probe, 1}(t)&=\frac{A_0}{\sqrt \pi a}\left(1-2\frac{(t-t_p)^2}{a^2}\right)e^{-(t-t_p)^2/a^2}\\
A_{\rm probe, 1}(t)&=-\frac{A_0}{\sqrt \pi}\frac{t-t_p}{a}e^{-(t-t_p)^2/a^2},
\end{align}
with the peak electric field $E_0=A_0/\sqrt{\pi}a$.

\begin{figure}[t]
\centering
\includegraphics[width=\columnwidth]{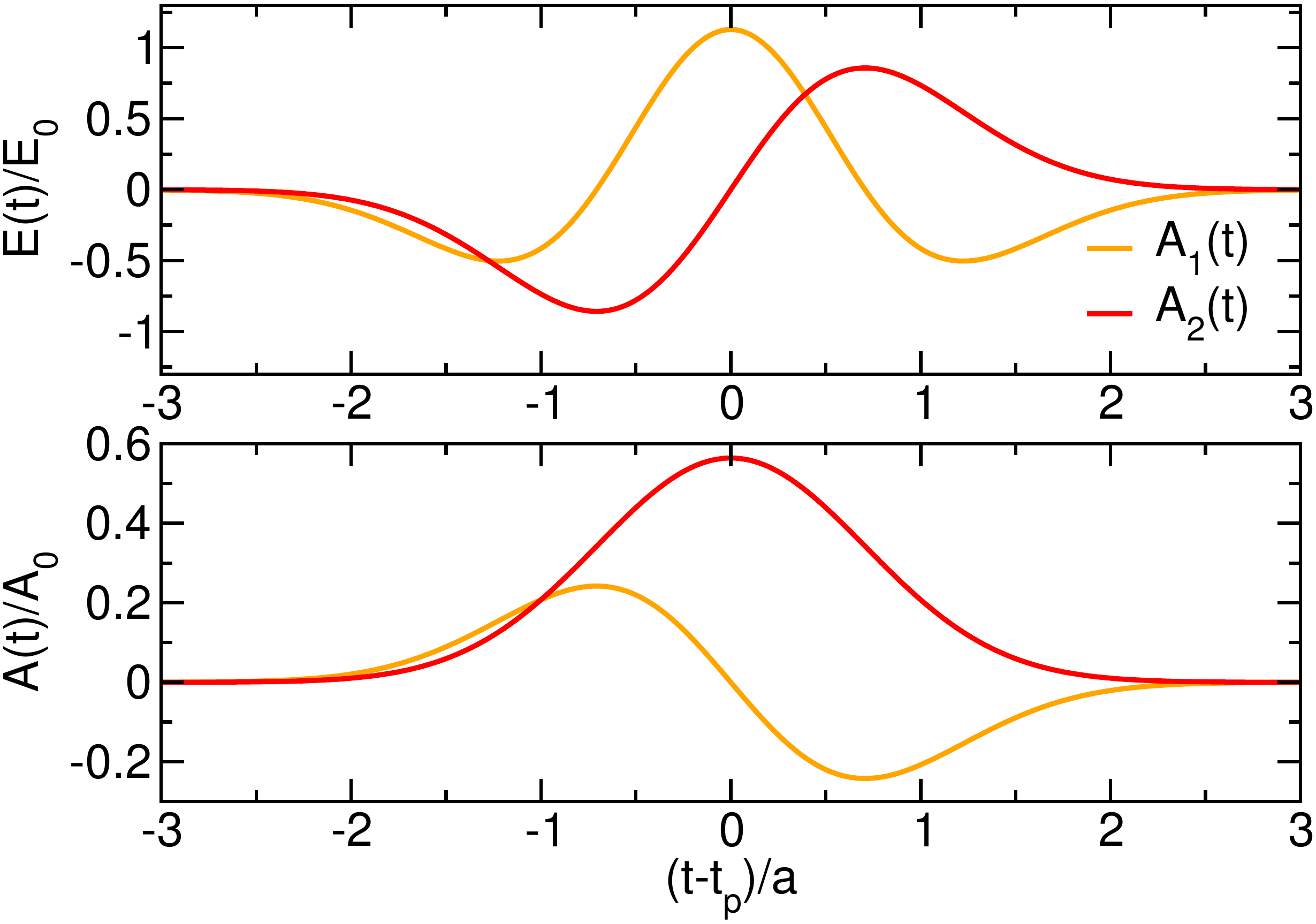}
\caption{Top panel: Electric field profile for probe functions given by the functional form of the vector potentials $A_{\rm probe, 1}(t)$ or $A_{\rm probe, 2}(t)$ with $A(t)=-\partial_t E(t)$. Bottom panel: corresponding vector potential. }
\label{fig:funcft}
\end{figure}

Here we propose a second form of the probe field, which will allow for a particularly simple reconstruction of the time dependent superfluid stiffness (the hallmark of a superconductor)   
\begin{align}
E_{\rm probe, 2}(t)&=\frac{2A_0}{\sqrt \pi a}\frac{(t-t_p)}{a}e^{-(t-t_p)^2/a^2}\\
A_{\rm probe, 2}(t)&=\frac{A_0}{\sqrt \pi }e^{-(t-t_p)^2/a^2}.
\end{align}
The parameter $a$ tunes the width of the probe pulses. In the following we will denote probe-pulses following the functional form $A_1(t)$ or $A_2(t)$ as type-I and type-II probe pulses, respectively. These functional forms are depicted in Fig.~\ref{fig:funcft}.

\begin{figure}
\begin{center} 
\includegraphics[width=\columnwidth, angle=-0]{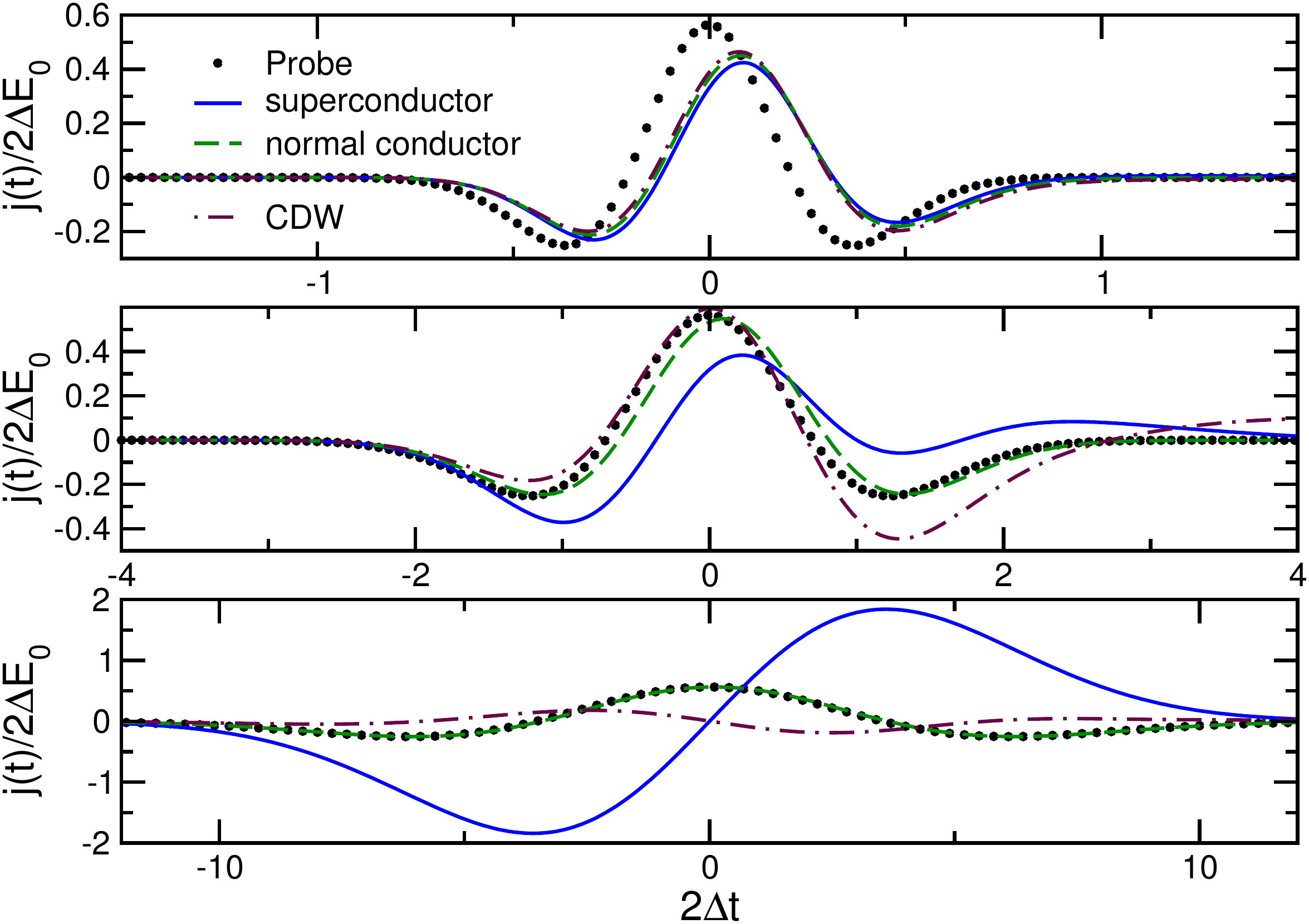}
\caption{Normal state (dashed lines), superconducting state (solid lines $A=1$) and charge density wave state (dashed-dotted lines) currents produced by the electric field pulse $E_1$ (dotted lines) for three pulse widths: $2\Delta a=0.3$ (top panel), $2\Delta a=1.0$ (central panel), $2\Delta a=5$ (lower panel). Note that different panels display different time ranges.}  
\label{simulateexptA1}
\end{center}
\end{figure}

Fig.~\ref{simulateexptA1} shows calculated superconducting, charge density and normal state currents for constant gap  for a type-I probe pulse and for three cases: a narrow pulse (width parameter $2\Delta a=0.3$, small compared to the inverse of the  gap), a pulse of width comparable to the inverse gap ($2\Delta a =1$) and a pulse of width much greater than the inverse gap ($2\Delta a=5$). The  superconducting, normal conducting and charge density wave currents are shown along with the pulse profile. In the short pulse case all three responses are almost indistinguishable because the current changes sign before the electronic response has time to build up, and the responses lag the pulse by a time set by the inverse of the scattering rate $\gamma/2\Delta$, chosen here to be 10. In the intermediate and longer pulse cases we see that the electric field varies slowly enough that the normal-state current essentially follows the field profile, while the superconducting state current is different, reflecting the supercurrent effects, which now have some time to develop. In the charge density wave we find a suppression of the current and the initial response of the current is of opposite sign compared to the electric field pulse.  We see that to obtain a significant difference between the normal and gapped currents we must use a pulse with a width which is at least of the order of the superconducting gap if not larger. 

\begin{figure}
\begin{center} 
\includegraphics[width=\columnwidth, angle=-0]{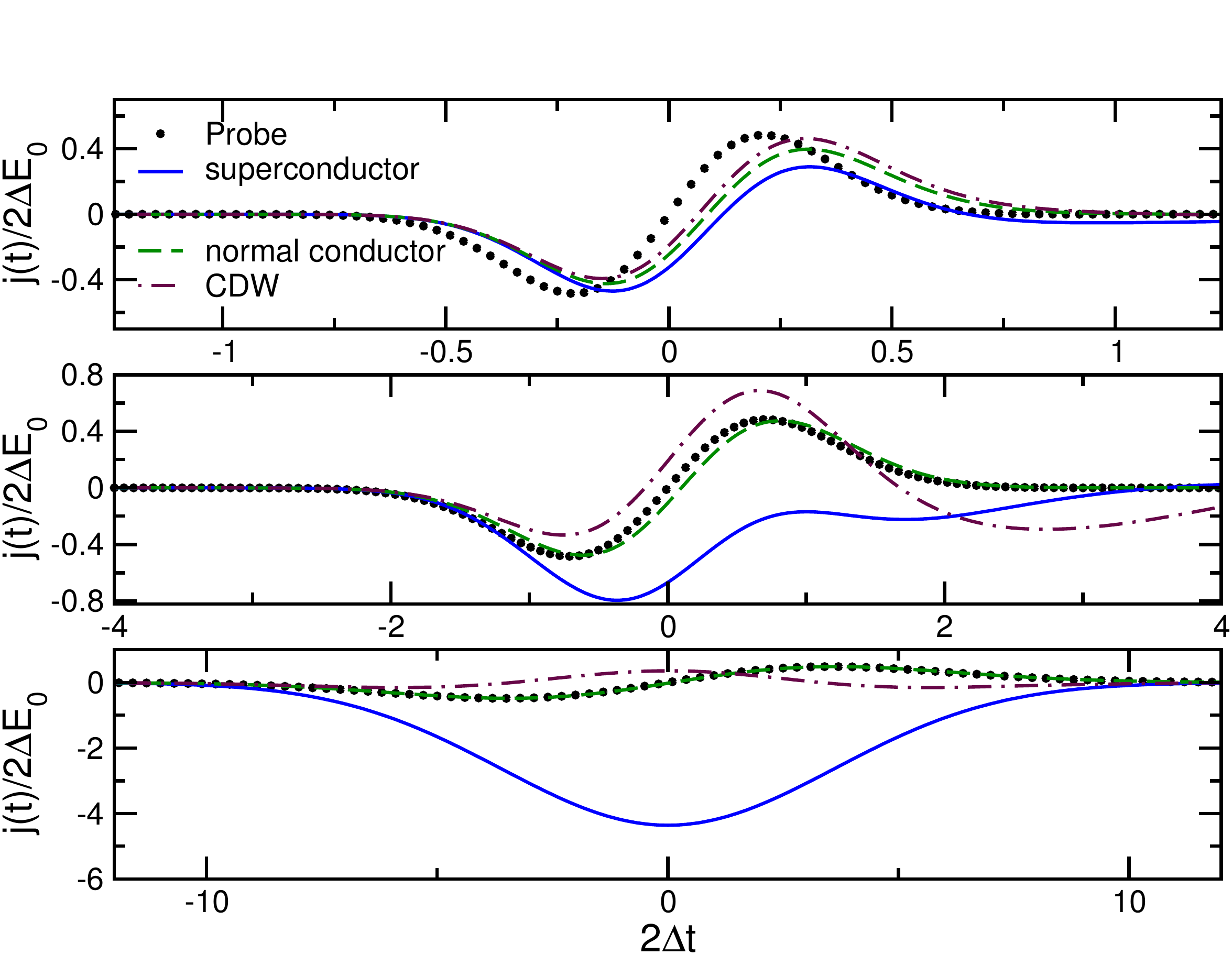}
\caption{Normal state (dashed lines), superconducting state (solid lines $A=1$)  and charge density wave state (dashed-dotted lines) currents produced by the electric field pulse $E_2$ (dotted lines) for three pulse widths: $2\Delta a=0.3$ (top panel), $2\Delta a=1.0$ (central panel), $2\Delta a=5$ (lower panel). Note that different panels display different time ranges. }  
\label{simulateexptA2}
\end{center}
\end{figure}

\begin{figure}
\begin{center} 
\includegraphics[width=\columnwidth, angle=-0]{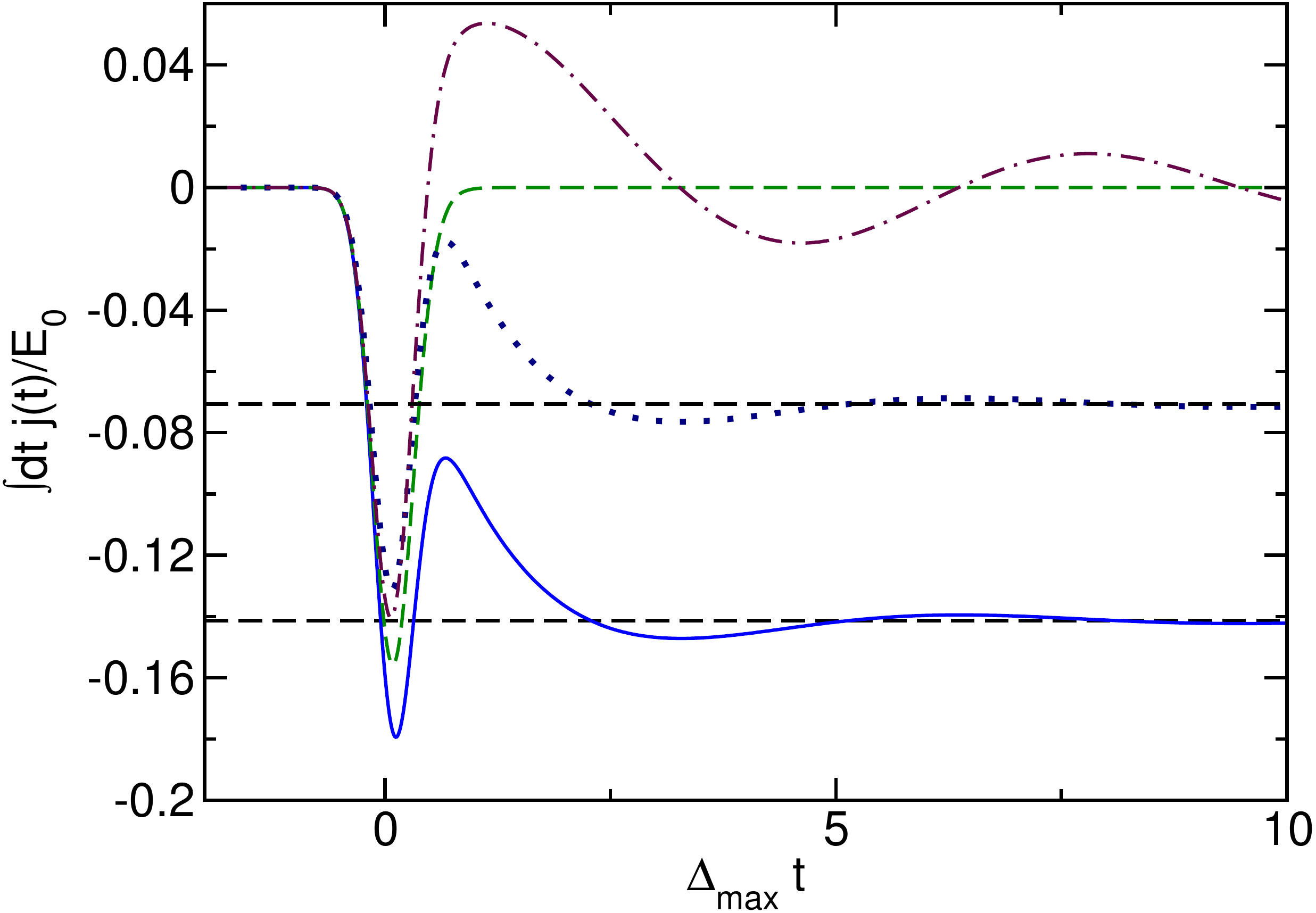}
\caption{Normal state (dashed line), superconducting state (solid line $A=1$ and dotted line $A=0.5$) and charge density wave state (dashed-dotted line) integrated currents of the currents shown in the upper panel of Fig.~\ref{simulateexptA2}.
For $t\to\infty$ the integrated currents are proportional to the superfluid stiffness (zero in normal and charge density wave state, $\sim A\Delta$ in superconducting state). The dashed horizontal lines indicates $-\rho_S a^2$ for $A=0.5$ and $A=1$. 
}  
\label{simulateexptA2_2}
\end{center}
\end{figure}

Fig.~\ref{simulateexptA2} shows the same as Fig.~\ref{simulateexptA1}, but for a type-II probe pulse. The behavior in time space is very similar to the conclusions drawn for the type-I case, however the integrated current reveals an interesting property in the superconducting case. This behavior is shown in Fig.~\ref{simulateexptA2_2} for $2\Delta a=0.3$.  Since the electric field is first negative and then positive a finite supercurrent flows in the superconducting state for a  time given by the width $a$, which sums up to a negative contribution. The contribution to the integrated current coming from the higher frequency part of the conductivity exactly cancel in the short pulse limit. Therefore, concentrating on small $a$ the normal and charge density wave state integrated currents approach zero asymptotically which is in clear contrast to the superconductor, where the non-zero supercurrent gives a small contribution $\int dt \; j(t)=-\rho_S a^2$ in the short pulse limit. Additionally to the superconducting $A=1$, normal and charge density wave state result we show the results for a superconducting state where $A=0.5$. The integrated current in this scenario is indeed halved indicating that the integrated current really probes the stiffness only. This will guide our intuition in the non-equilibrium case. If the gap profile and with it the superfluid stiffness does not evolve too quickly, it should be possible to reconstruct its value along the same lines as outlined in the previous paragraph using the integrated current in a type-II probe setup.

\section{Non-equilibrium Conductivity: Theory \label{details}}
In this section we will present the calculation of the non-equilibrium optical conductivity of a dirty superconductor exhibiting a time dependent gap $\Delta(t)$,\cite{footnote1} which is an input to the theory (the fully self-consistent treatment of transient order is left to another paper). We  follow the  strategy first used by Mattis and Bardeen \cite{Mattis58} to calculate the equilibrium dirty-limit conductivity by calculating the clean limit conductivity as a function of momentum and then averaging over momentum. This procedure was recently used by Chou, Liao  and Foster to obtain the conductivity at long times after a quench.\cite{Chou16} 

We study  a BCS-type s-wave superconductor
\begin{equation}
H=\sum\limits_{k,\sigma=\uparrow\downarrow} \epsilon_k c_{k,\sigma}^\dagger c_{k,\sigma} + \Delta(t) \sum\limits_k  c^\dagger_{k,\uparrow} c^\dagger_{-k,\downarrow}+{\rm H.c.},\label{eq:Hini}
\end{equation}
with time dependent gap $\Delta(t)$. Here  $c^{(\dagger)}_{k,\sigma}$ annihilates (creates) a fermion in the single particle state characterized by momentum $k$ and spin $\sigma$. We use the language of Nambu-vectors $\Psi_k^\dagger=(c^\dagger_{k,\uparrow},c_{-k,\downarrow})$ to rewrite the Hamiltonian as  
\begin{equation}
H=\sum\limits_{k} \Psi^\dagger_k \begin{pmatrix}
\epsilon_k&\Delta(t)\\
\Delta(t)^*&-\epsilon_k
\end{pmatrix}\Psi_k.
\label{eq:Hsc}
\end{equation}
Similarly we can describe a perfectly nested charge density wave (with ordering vector $Q$) by one minor change, i.e. the Nambu spinor reads $\Psi_k^\dagger=(c^\dagger_{k,\uparrow},c^\dagger_{k+Q,\uparrow})$. The Hamiltonian then takes the same form as Eq.~\eqref{eq:Hsc}.

\begin{figure}[t]
\centering
\includegraphics[width=\columnwidth]{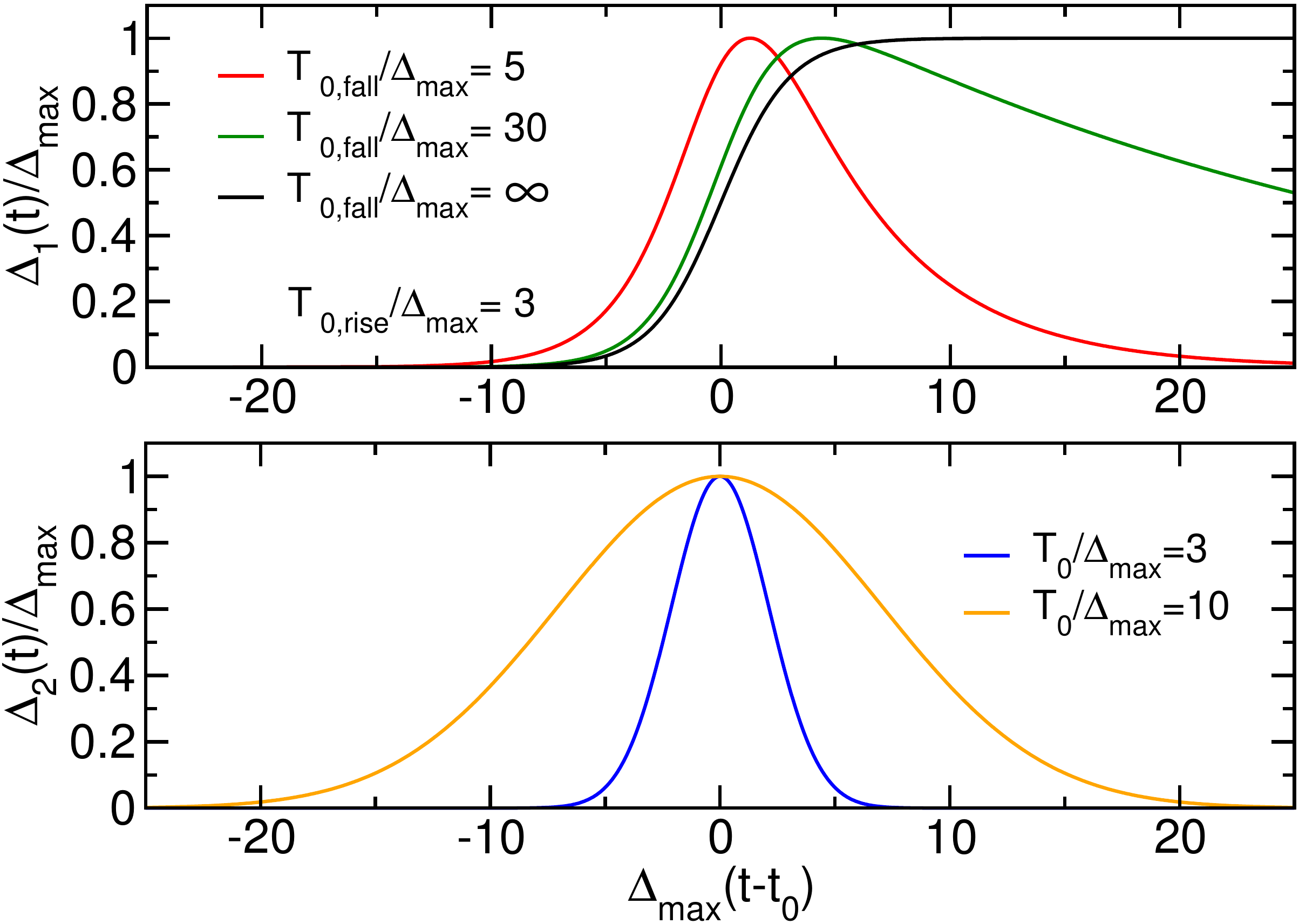}
\caption{Gap profiles for $\Delta_1(t)$ Eq.~\eqref{eq:Delta1} (top panel) as well as $\Delta_2(t)$ Eq.~\eqref{eq:Delta2} (bottom panel) with $\Delta_{\rm ini}/\Delta_{\rm max}=0$, $\Delta_{\rm max}t_0=0$ and $\Delta_{\rm max} T_{0(,{\rm rise/fall})}$ given in the legend.}
\label{fig:funcDt}
\end{figure}

In a non-equilibrium superconductor or charge density wave the time dependence of the gap is determined by the time dependence of a pump field  and the physics of the system in question. Here, we  focus on the observable consequences of this time dependence. We therefore focus on two very general gap  profiles, representative of those discussed in the theoretical literature \cite{Quenchlit,barankov2004collective,Kennes16,Sentef16,Chou16}; this renders our results independent of the details of how precisely the gap is induced):
\begin{align}
\Delta_1(t)=&\Delta_{\rm ini}+ \frac{C}{2}\left(\Delta_{\rm max}-\Delta_{\rm ini}\right)\left(\tanh\left(\frac{t-t_0}{T_{0,{\rm rise}}}\right)+1\right)\notag\\
&\phantom{\Delta_{\rm ini}+}\times e^{-t/T_{0,{\rm fall}}},\label{eq:Delta1}\\
\Delta_2(t)=& \Delta_{\rm ini}+ \left(\Delta_{\rm max}-\Delta_{\rm ini}\right)e^{-\left(\frac{t-t_0}{T_0}\right)^2},\label{eq:Delta2}
\end{align}
where the constant $C$ is chosen such that the gap rises from $\Delta_{\rm ini}$ to $\Delta_{\rm max}$ (and subsequently settles back down to $\Delta_{\rm ini}$ for finite $T_{0,{\rm fall}}$). In these protocols $t_0$ describes the delay time, while $\Delta_{\rm ini}$ and $\Delta_{\rm max}$ are the initial and maximum value of the gap. $\Delta_1(t)$ allows for an asymmetric rise and fall time of the time dependent gap given by $T_{0,{\rm rise}}$ and $T_{0,{\rm fall}}$, respectively. In contrast  $\Delta_2(t)$ exhibits a symmetric rise and fall time $T_0$ of the gap function. Their functional forms are depicted in Fig.~\ref{fig:funcDt}. For a given $\Delta(t)$ one can determine the non-equilibrium Green's functions $G^{\rm R/K/A}(k,t,t')$ as shown in the appendix.  We choose the two gap profiles of Eqs.~\eqref{eq:Delta1} and \eqref{eq:Delta2} these are  representative of the  classes of  time  dependent situations that have been discussed in the literature and in particular  
are consistent with the results reported for the self-consistent solution of the gap equation in Ref.~\onlinecite{Kennes16}. We note that in condensed matter systems and in many theories such as that of Ref.~\onlinecite{Kennes16} the electrons can exchange energy with a reservoir. Without this thermalization mechanism the self-consistent solution would show long-lived large amplitude oscillations due to the integrability of the system.\cite{barankov2004collective} Gap profiles including these long-lived oscillations  can be treated in the same formalism as presented here, but are not studied here as we expect that such oscillations are washed out in condensed matter systems by the coupling of the electrons to the environment. While the time-dependent BCS approximation used here is sufficient for our purposes of understanding the qualitative aspects of the transient response of superconducting and density wave states, it omits inelastic effects caused by the nonequilibrium drive, which may lead to some broadening of the results.

Given $G$ we now write the paramagnetic term in the current-current correlator $\chi$ (Eq.~\eqref{chidef1} but with momentum labels restored) as 
\begin{eqnarray}
\chi_{JJ}^{para}(q;t,t^\prime)&=&\frac{-i}{2}Tr\bigg[\hat{J}^{kq}_{op}G^R(k+q;t,t^\prime)\hat{J}^{q k}_{op}G^K(k;t^\prime,t)
\nonumber \\
&&+ \hat{J}^{kq}_{op}G^K(k+q;t,t^\prime)\hat{J}^{q k}_{op}G^A(k;t^\prime,t)\bigg]
\label{chidef2}
\end{eqnarray}
where the trace is over momentum ($k$), spin and Nambu indices. Mattis and Bardeen\cite{Mattis58} observed that in the presence of scattering leading to a mean free path $l$ one may average $\chi$ over $q\sim l^{-1}$. In the dirty limit, $l$ is much less than the bare superconducting coherence length $\xi_0$ or equivalently the scattering rate $\gamma\gg\Delta$ and we also assume that the mean free path is longer than an inverse Fermi wavevector ($k_Fl>1$) so that the eigenstates are not localized. The difference $\chi_S-\chi_N$ between the superconducting and normal state $\chi$ is dominated by momenta within a $x_0^{-1}$ of the Fermi surface (energies within $\Delta$ of the chemical potential). In this limit, we may then integrate independently over the magnitudes of $\varepsilon_k$ and $\varepsilon_{k+q}$, perform the angular integrals as in the usual conductivity calculation (the condition $k_Fl>1$ means that we may neglect any variation of the current operator except the direction), collect the prefactors into those corresponding to the normal state conductivity and add back the normal state contribution, obtaining 
\begin{eqnarray}
\chi_{JJ}^{para}(t,t^\prime)&=&\frac{K}{\gamma}\frac{-i}{2}Tr\bigg[\tau_3G^R(t,t^\prime)\tau_3G^K(t^\prime,t)
\nonumber \\
&&+ \tau_3G^K(t,t^\prime)\tau_3G^A(t^\prime,t)\bigg],
\label{chidef3}
\end{eqnarray}
where the $\tau_3$ are matrices acting in Nambu space representing the type II coherence factors of the optical process and the Green functions without the momentum indices represent the results of integration over the energy variable. 
Equation~\eqref{chidef3} expresses the response of the current in terms of the gap $\Delta(t)$ dependent Green's functions as well as the vector potential $A(t)$.  


The correlator defined in Eq.~\eqref{chidef3} needs regularization at short relative times $t'=t^-$. In our calculation the regularization is provided by the finite bandwidth. The key insight of Mattis and Bardeen is that on physically relevant time scales, e.g. $\Delta_{\rm max}^{-1}$, results are independent of the regularization up to a scale factor, which is set by the f-sum rule Eq.~\eqref{Kdef1}, $\sigma(t,t)=K(t)$. In most of the results presented below we subtract out the non-universal short-time behavior by working with $\Delta\chi(t,t')=\chi_{JJ}^{para}(t,t^\prime,\Delta(t))-\chi_{JJ}^{para}(t,t^\prime,\Delta(t)=0)$ and dividing by a suitable quantity.

\section{Non-equilibrium Conductivity: Results in time space \label{resultstime}}

\subsection{Introduciton}
\begin{figure}[t]
\centering
\includegraphics[width=\columnwidth]{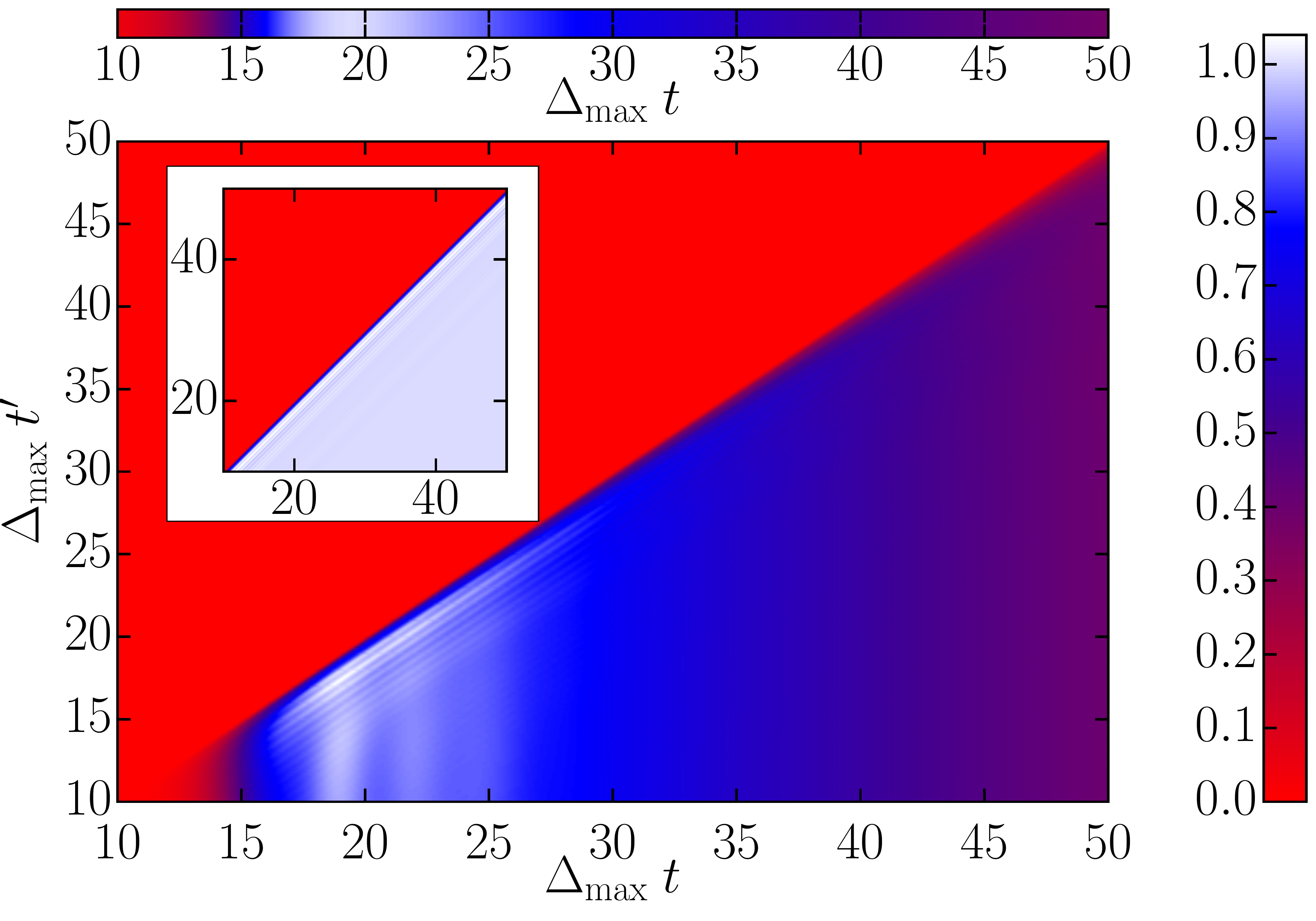}
\caption{Main panel: Two-time dependent conductivity $[\sigma^{\rm s}(t,t')-\sigma^{\rm n}(t,t')]/\sigma_0$  for $\Delta(t)=\Delta_1(t)$ with $\Delta_{\rm ini}/\Delta_{\rm max}=0$, $\Delta_{\rm max}t_0=25$, $\Delta_{\rm max}T_{0,{\rm rise}}=3$ and  $\Delta_{\rm max}T_{0,{\rm fall}}=30$ for a superconducting state, with $\sigma_0=\lim\limits_{t\to\infty}\sigma^{\Delta(t)=\Delta_{\rm max}}(t,0)$. The small panel above the main panel shows the time evolution of $\Delta(t)$ in the same color scheme. The inset gives the equilibrium optical conductivity at the maximum value of the gap $\Delta=\Delta_{\rm max}$. Due to time translation invariance there are only diagonal features in the inset. }
\label{fig:twotime}
\end{figure}

Within the above framework $\sigma(t,t')$ can be determined for the normal, superconducting or charge density wave state case.  This (two-times) function is the fundamental linear response object of interest, which determines the physics unambiguously. Figure~\ref{fig:twotime} shows exemplary results for the two time dependent $\sigma(t,t')$ for $\Delta(t)=\Delta_1(t)$ with $\Delta_{\rm ini}/\Delta_{\rm max}=0$, $\Delta_{\rm max}t_0=25$, $\Delta_{\rm max}T_{0,{\rm rise}}=3$ and  $\Delta_{\rm max}T_{0,{\rm fall}}=30$ showing $[\sigma^{\rm s}(t,t')-\sigma^{\rm n}(t,t')]/\sigma_0$ for a superconducting state in the main panel, with $\sigma_0=\lim\limits_{t\to\infty}\sigma^{\Delta(t)=\Delta_{\rm max}}(t,0)$. We subtract the normal state conductivity to cancel out the Drude contribution and thus highlight only the superconducting part. For $t'>t$ we find zero as expected. For $t>t'$ the main panel shows that $\sigma(t,t')$ quickly follows the instantaneous gap $\Delta(t)$, which is depicted in the small panel above the main panel, superimposed with small transient oscillations following the probe field.   The inset shows the equilibrium optical conductivity at the maximum value of the gap $\Delta=\Delta_{\rm max}$. In contrast to the equilibrium optical conductivity where only diagonal features in $t$ and $t'$ occur (due to time translation invariance), the non-equilibrium optical conductivity shows two types of features. Those along the diagonal resembling the adiabatic (equilibrium) physics and those along the horizontal reflecting the time dependent change of the gap function. Comparing the color coding of the main panel to the small panel above, one finds that overall the instantaneous value of the gap function $\Delta(t)$ (proportional to the stiffness in equilibrium within our BCS approach) can be read off with reasonable precision from the two-times conductivity by considering a given value of $t$ not too close to $t'$. 

However, reconstruction of the full two-times function $\sigma(t,t')$ would require an exhaustive number of measurements in experiments as one needs to sample a two dimensional $(t,t')$ grid. Therefore, now we  consider the simpler question of what can be learned about the optical conductivity if the incident light-probe profile and consequently $E(t)$ is fixed. In this section we report on the time domain current response $j(t)$. Here we show how engineering optimal probe pulse shapes can reveal the underlying non-equilibrium physics most easily. 


\subsection{Results in Time Space}

\begin{figure}[t]
\centering
\includegraphics[width=\columnwidth]{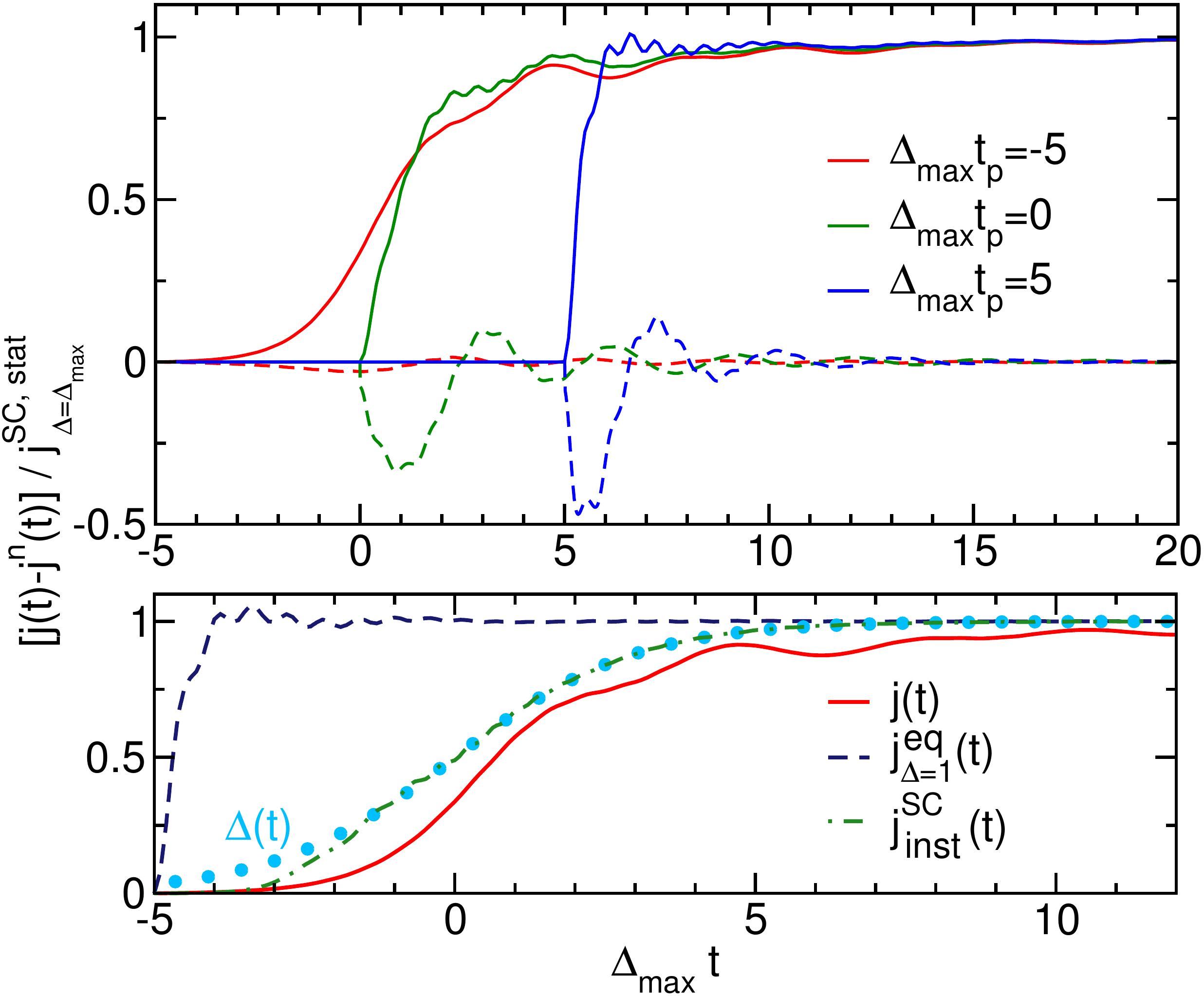}
\caption{Top panel: Response of the current $j(t)$ (measured relative to normal state current $j^{\rm n}(t)$ and normalized to the equilibrium supercurrent value at the maximum value of the gap $\Delta_{\rm max}$) to an electric field pulse $E(t)\sim \delta(t-t_p)$ for quenches into the superconducting state (solid lines) and into the charge density wave state (dashed lines) for a gap profile (shown as blue dots in the bottom panel) $\Delta_1(t)$ with $\Delta_{\rm ini}/\Delta_{\rm max}=0$, $\Delta_{\rm max}t_0=0$ and $\Delta_{\rm max}T_{0,{\rm rise}}=3$ (quench means $T_{0,{\rm fall}}\to \infty$). The electric probe pulse is applied at times $\Delta_{\rm max}t_p=-5,0,5$ (red, green, blue line) relative to the half way point of the gap rise.  Bottom panel: solid line (red online) expanded view of current for $\Delta_{\rm max}t_p=-5$ from main panel; dotted line (blue online) gap profile $\Delta(t)$, dashed line (maroon online) time response for equilibrium superconducting state with $\Delta=\Delta_{\rm max}$, dashed dotted line (green online) instantaneous approximation to the current described in main text.}  
\label{fig:Current_Epulse}
\end{figure}

\begin{figure}[t]
\centering
\includegraphics[width=\columnwidth]{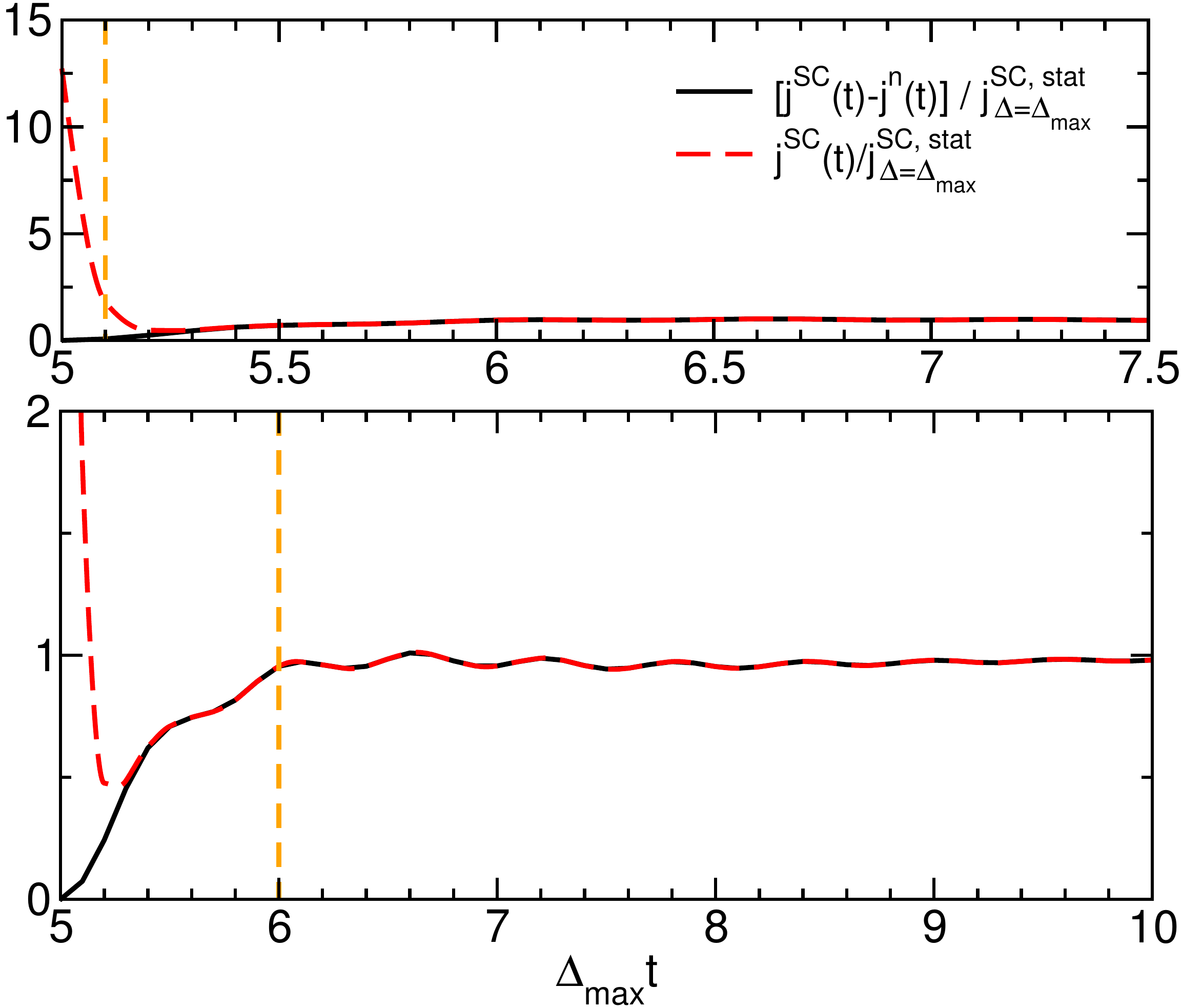}
\caption{Top panel: solid line (black online) superconducitng state data of  Fig.~\ref{fig:Current_Epulse} for  $\Delta_{\rm max}t_p=5$. Heavy dashed line (red online)  total current including short time contribution obtained from a physical regularization via a scattering rate $\gamma/2\Delta_{\rm max}=10$. Vertical thin dashed line (orange online) shows the time scale $\sim 1/\gamma$. Bottom panel: data of the upper panel replotted for different scales for the y- and x-axes. Vertical line (orange online) indicates the time scale $\sim 1/\Delta_{\rm max}$.  }  
\label{fig:Current_Epulse3}
\end{figure}

\begin{figure}[t]
\centering
\includegraphics[width=\columnwidth]{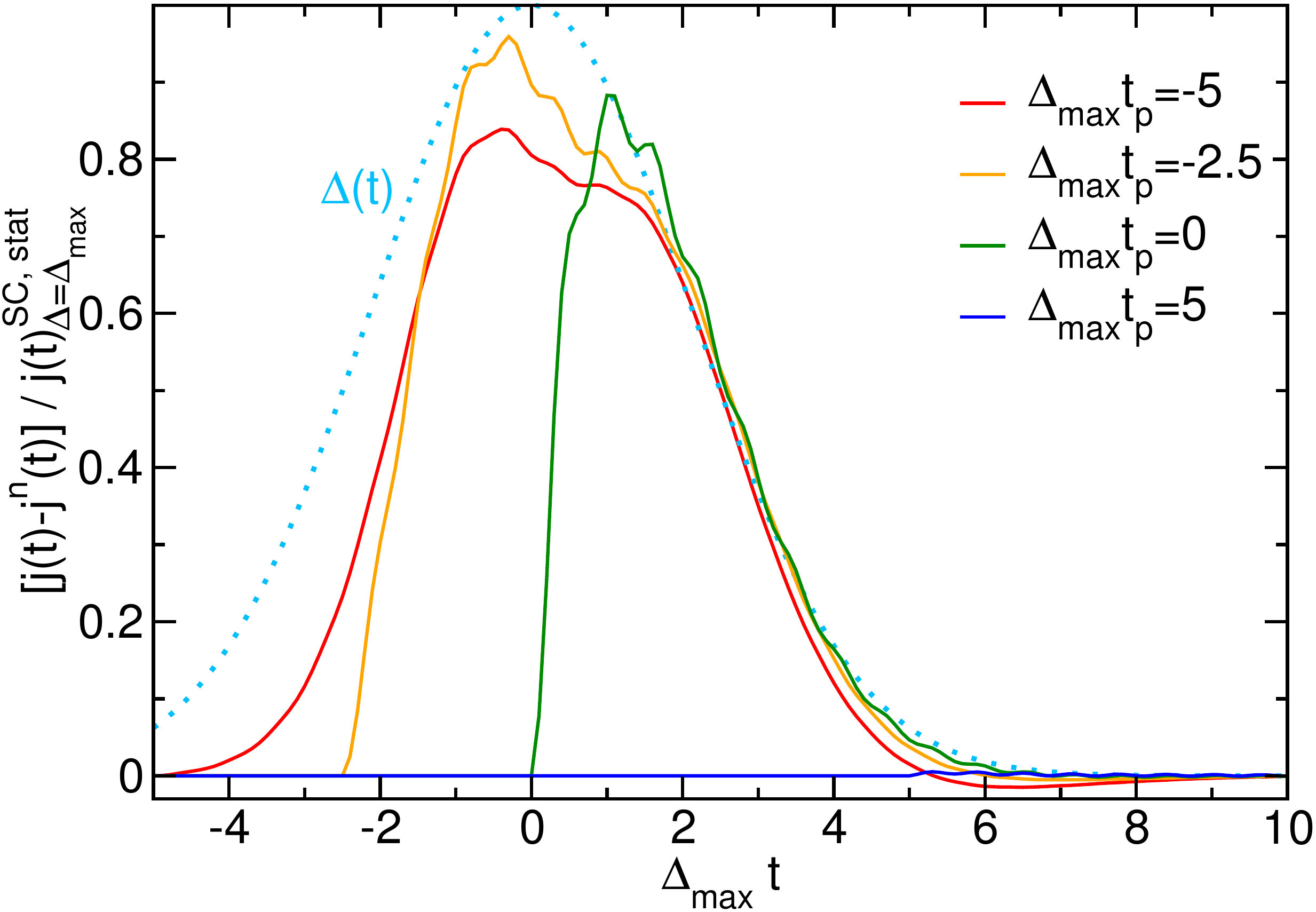}
\caption{Same as Fig.~\ref{fig:Current_Epulse}, but for $\Delta_2(t)$ (shown as blue dots) with  $\Delta_{\rm ini}/\Delta_{\rm max}=0$, $\Delta_{\rm max}t_0=0$ and $\Delta_{\rm max}T_{0}=3$ and concentrating on the superconducting state. The electric probe pulse is applied at $\Delta_{\rm max}t_p=-5,-2.5,0,5$ (red,orange,  green, blue line).}
\label{fig:Current_Epulse2}
\end{figure}

We start our discussion with the simplest case, of a delta function $E(t)=E_0\delta(t-t_p)$ pulse, which although experimentally very difficult to implement, provides interesting insights useful for the subsequent discussion of the experimentally relevant pulses given below. For the  delta function $E(t)=E_0\delta(t-t_p)$ pulse the current $j(t) $ is identical to $\sigma(t,t_p)$.
We note one important point: our calculation assumes that by the time the pulse is applied the superfluid stiffness is large enough that our initial gauge choice of phase $\phi=0$ is not affected by the probe, so that the superfluid response is directly proportional to the vector potential. If the probe is applied  at very early times, before the superfluid stiffness is appreciable, then the final superconducting state will be described by a phase $\phi\neq 0$, $\nabla \phi\neq 0$ and the expression for the supercurrent would be different. This case will be discussed in a subsequent paper  (see also footnote \onlinecite{footnote1}). 

In the top panel of Fig.~\ref{fig:Current_Epulse} we summarize our results for a gap that is turned on from $\Delta_{\rm ini}/\Delta_{\rm max}=0$ to $\Delta_{\rm max}$ on a time scale of $\Delta_{\rm max}T_{0,{\rm rise}}=3$ and never decays back to zero ($\Delta_{\rm max}T_{0,{\rm fall}}\to\infty$).  We apply a probe pulse at different $t_p$ and show the current difference between the superconducting and normal state or charge density wave state and normal state, respectively. We find that the asymptotic value of the current after long times is given by the corresponding equilibrium value. If the pulse is applied in the distant future with respect to the turning on of the gap the time evolution looks the same as the equilibrium one. As in equilibrium (see above) the charge density wave state and superconducting state behave opposite in the short time limit, meaning that the first rises up, while the latter become negative first. The time scales of rise times and oscillations are very roughly given by the instantaneous $\sim 1/\Delta(t)$. 

The bottom panel shows an expanded view of the non-equilibrium current for one particular case $\Delta_{\rm max}t_p=-5$. We compare this to two other cases. First, we show as the dashed line the equilibrium result at maximum gap. This result is characterized by a rapid turn on of the superconducting current on a time scale   set by the gap. Second, we show the "instantaneous" current obtained by rescaling the equilibrium result to the instantaneous value of the gap $\Delta(t)$ and the time lag to $t\Delta(t)$. We see that this instantaneous current build more slowly than the equilibrium one, because the rise time is controlled by the instantaneous gap value, but by times $\sim-2$ joins the curve (blue dots) given by the long time limit of the current given by the instantaneous value of the gap. The actually non-equilibrium current lacks behind this curve indicating an even slower build up of the superfluid response and furthermore at long times converges slowly to the asymptotic value.

In Fig.~\ref{fig:Current_Epulse3} we show both the difference superconducting current plotted in Fig.~\ref{fig:Current_Epulse} for $\Delta_{\rm max}t_p=5$ and the  full current obtained by adding back the Drude regularization with a scattering rate $\gamma/2\Delta_{\rm max}=10$ (dashed red line). Depending on the relative value of the scattering rate and the gap, the initial rise of the supercurrent is masked by the initial decay of the drude current.

%

In Fig.~\ref{fig:Current_Epulse2} we show the same as in Fig.~\ref{fig:Current_Epulse}, but for a symmetrically rising and decaying gap function chosen by $\Delta_2(t)$ (shown as blue dots) with  $\Delta_{\rm ini}/\Delta_{\rm max}=0$, $\Delta_{\rm max}t_0=0$ and $\Delta_{\rm max}T_{0}=3$ and concentrating on the superconducting state. Again the rise times are roughly related to the instantaneous value of the gap $\sim1/\Delta(t)$. When $\Delta(t)$ decreases, the current closely follows the decrease in gap value with almost no time lag. Due to the lag of the non-equilibrium current described above the supercurrent never reaches its equilibrium value at $\Delta_{\rm max}$. Also there appears a small negative feature at the turn off of the gap if the probe is applied before the maximum of the gap function, which is beyond an instantaneous description. The negative feature decreases in height with the time distance between maximum of the gap function and the probe field becoming shorter.   

\begin{figure}[t]
\centering
\includegraphics[width=\columnwidth]{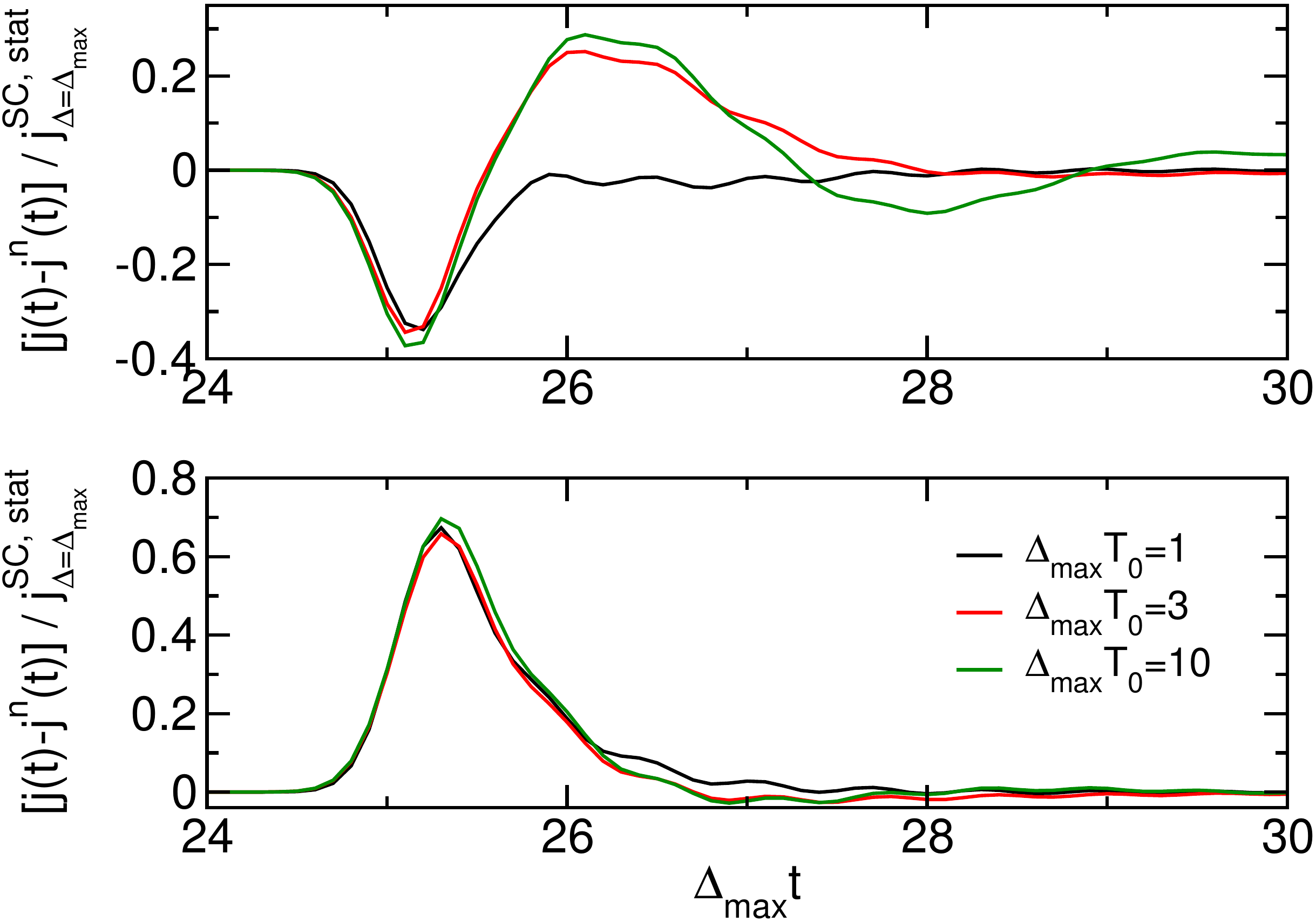}
\caption{ Real time currents after a type-II probe of width $\Delta_{\rm max}a=0.3$ for $\Delta(t)=\Delta_2(t)$ with $\Delta_{\rm ini}/\Delta_{\rm max}=0$, $\Delta_{\rm max}t_0=25$ and $\Delta_{\rm max}T_{0}=1$ (black lines), $\Delta_{\rm max}T_{0}=3$ (red lines) as well as  $\Delta_{\rm max}T_{0}=10$ (green line) for both the charge density wave state (upper panel) as well as the superconducting state (lower panel).}
\label{fig:transientneq}
\end{figure}

Now we turn to the experimentally more practical type-I and type-II probe pulses. 
We start by showing that the characteristic difference between a superconducting and a charge density wave state, i.e. the difference in the sign of the current after an electric field pulse identified above for the equilibrium case, holds even in strong non-equilibrium. This is summarized in Fig.~\ref{fig:transientneq} for the $\Delta(t)=\Delta_2(t)$ with $\Delta_{\rm ini}/\Delta_{\rm max}=0$, $\Delta_{\rm max}t_0=25$ and $\Delta_{\rm max}T_{0}=1$ (black lines), $\Delta_{\rm max}T_{0}=3$ (red lines) as well as  $\Delta_{\rm max}T_{0}=10$ (green line) for a a type-II probe pulse of width $\Delta_{\rm max}a=0.3$ (qualitatively similar results hold for the type-I pulse).  

\begin{figure}[t]
\centering
\includegraphics[width=\columnwidth]{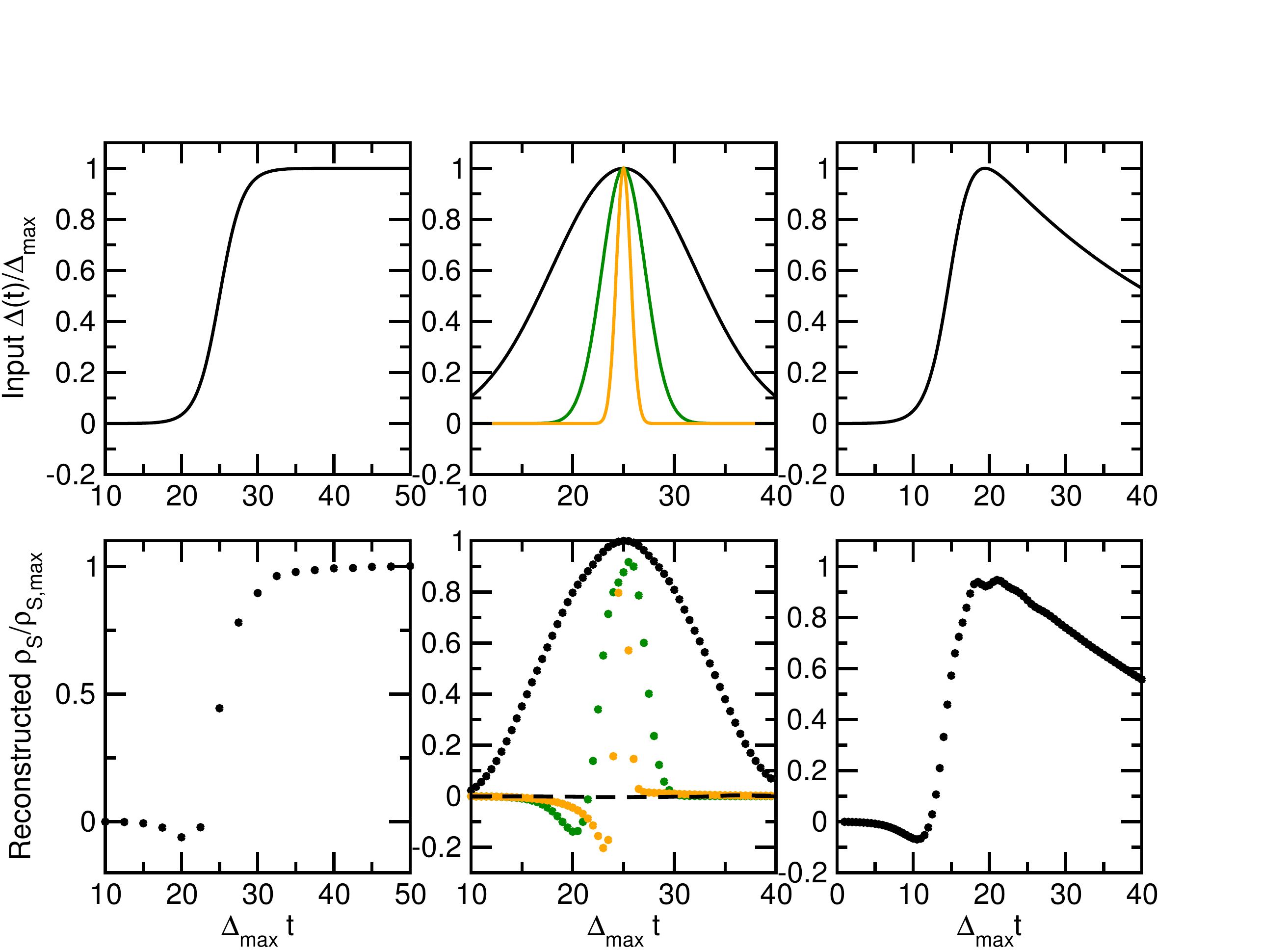}
\caption{Reconstructing the stiffness from the integrated current arising from a short type-II probe pulse with $\Delta_{\rm max}a=0.3$. Top panels: input form of $\Delta(t)$, which is $\Delta(t)=\Delta_1(t)$ with $\Delta_{\rm ini}/\Delta_{\rm max}=0$, $\Delta_{\rm max}t_0=0$, $\Delta_{\rm max}T_{0,{\rm rise}}=3$ and $\Delta_{\rm max}T_{0,{\rm fall}}\to \infty$ (left) or $\Delta_{\rm max}T_{0,{\rm fall}}\to 30$ (right) and $\Delta(t)=\Delta_2(t)$ (middle) with $\Delta_{\rm ini}/\Delta_{\rm max}=0$, $\Delta_{\rm max}t_0=25$ and $\Delta_{\rm max}T_{0}=1$ (orange lines), $\Delta_{\rm max}T_{0}=3$ (green lines) as well as  $\Delta_{\rm max}T_{0}=10$ (black line). Bottom panels: reconstruction of the stiffness via $\int dt\; j(t)= -a^2\rho_S$ of the corresponding upper panels. In the middle panel we plot also the reconstruction of the stiffness in a charge density wave state with the same time dependent gap $\Delta(t)$, which is as expected zero (dashed line).}
\label{fig:Reconstruct}
\end{figure}

Next we show how engineering type-II probe pulses of finite width can be useful to reconstruct the stiffness from the integrated current directly by using $\int dt\; j(t)= -a^2\rho_S$. This was demonstrated for equilibrium in Fig.~\ref{simulateexptA2}. We consider different input profiles of the gap function $\Delta(t)$ in Fig.~\ref{fig:Reconstruct}.
We find that for short pulses ($\Delta_{\rm max} a=0.3$) the reconstruction works very well as long as the gap does change slowly, but if the gap feature a more rapid evolution the reconstruction gets increasingly worse. The reconstructed stiffness at a sharply rising gap even shows a negative feature, which becomes more prominent the quicker the gap is varied. This negative feature can be understood from the full non-equilibrium current lagging behind the instantaneous approximation introduced above. As the gap is turned on rapidly the contribution to the integrated current from the higher frequency part of the conductivity of the first (negative) half of the electric field pulse  does not cancel exactly the second (positive) half  and leaves behind a small positive imbalance in the integrated current. If the probe pulse is short enough the integrated supercurrent $= -a^2\rho_S$ is small and can be overcompensated by the former contribution to the integrated current. Using the (in this case false) relation $\int dt\; j(t)= -a^2\rho_S$ then leads to a negative reconstruction of the stiffness.

\begin{figure}[t]
\centering
\includegraphics[width=\columnwidth]{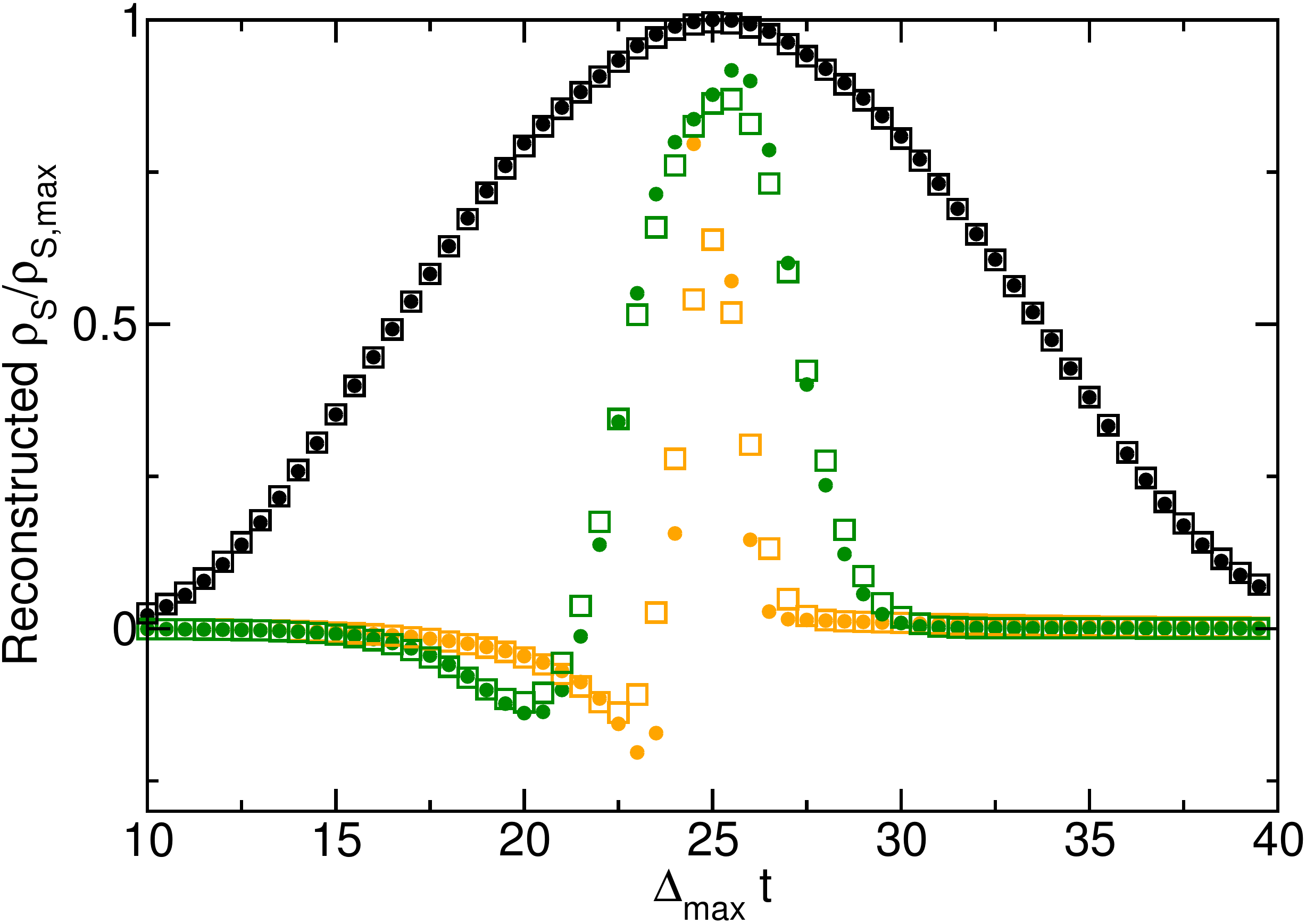}
\caption{The same as Fig.~\ref{fig:Reconstruct}, but showing only the reconstruction of the stiffness for the middle panels of Fig.~\ref{fig:Reconstruct}, so $\Delta(t)=\Delta_2(t)$ with  $\Delta_{\rm ini}/\Delta_{\rm max}=0$, $\Delta_{\rm max}t_0=25$ and $\Delta_{\rm max}T_{0}=1$ (orange symbols), $\Delta_{\rm max}T_{0}=3$ (green symbols) as well as  $\Delta_{\rm max}T_{0}=10$ (black symbols). We compare two width's of the type-II probe pulse $\Delta_{\rm max}a=0.3$ (filled circles) and $\Delta_{\rm max}a=1$ (open squares).  }
\label{fig:Reconstruct2}
\end{figure}

In Fig.~\ref{fig:Reconstruct2} we compare the reconstruction of the stiffness obtained in the same way as for Fig.~\ref{fig:Reconstruct} via the integrated current, but comparing $\Delta_{\rm max}a=0.3$ to a broader probe pulse $\Delta_{\rm max} a=1.0$. We concentrate on the $\Delta(t)$ analyzed in the middle panels of Fig.~\ref{fig:Reconstruct2}, so $\Delta(t)=\Delta_2(t)$ with $\Delta_{\rm ini}/\Delta_{\rm max}=0$, $\Delta_{\rm max}t_0=25$ and $\Delta_{\rm max}T_{0}=1$ (orange symbols), $\Delta_{\rm max}T_{0}=3$ (green symbols) as well as  $\Delta_{\rm max}T_{0}=10$ (black symbols). If the width of the time dependent gap is large compared to the probe's width the reconstructions give identical results. As the width of the probe pulse approaches the width of $\Delta(t)$ two effect show up. First, the wrongly reconstructed negative stiffness is reduced (as the supercurrent contribution mentioned above becomes larger). Second,  the probe pulse averages over an increasingly large time range of $\Delta(t)$. This is most clearly reflected in the maximum value of the reconstruction, which is reduced for larger pulse width.

\section{Non-equilibrium Conductivity: Results in Frequency Space \label{resultsfreq}}
In this section we turn to the frequency domain. In the nonequilibrium situation the conductivity is  a function of two frequencies, but what is often presented is a function of a single frequency defined as the ratio between the measured current at a given frequency and the applied field at the same frequency, $``\sigma(\omega)''=j(\omega)/E(\omega)$ (note that in experiments the introduction of variable delays between pump and probe pulses\cite{Larsen11,Kindt99,Averitt00} can provide more information). By considering the frequency-domain response to the  experimentally motivated type-I and type-II probe pulses we study the extent to which the single-frequency conductivity provides a clear representation of the physics.   
\subsection{Short Probe-Pulses}
It is convenient to first consider the limit $a\to 0$. This limit corresponds to a delta-function $A(t)\sim \delta(t-t_p)$ (type II pulse) or  derivative of a delta-function $A(t)\sim \delta'(t-t_p)$  (type I pulse) of the vector potential. This limit is particularly interesting as from the delta-distribution $A(t)$ pulse in principal the results for a general form of $A(t)$ can be generated, and the variable time delay tricks \cite{Larsen11,Kindt99,Averitt00} can in effect reveal the response associated with the very narrow pulse limit.

One of the hallmarks of a superconductor is an infinitely lived super current as a response to an electric field pulse. In equilibrium the signature of this current can be found in the imaginary part of the optical conductivity, which diverges as $p\Delta/\omega$ at small $\omega$  where $p$ is some  prefactor that depends on the details of the dispersion $\epsilon_k$ of the system only.  Of course a true infinitely lived supercurrent is not accessible in a transiently lived superconducting state (even if the gap evolves adiabatically).  In the optical conductivity, if probed by an electric field pulse $E=E_0\delta(t-t_p)$, of a transient superconductor with life time $T_0$ this reflects in a downturn of the apparent $1/\omega$ divergence back to zero at a frequency scale $\omega\sim 1/T_0$.\cite{Kennes16}

However, in experiments the probe pulses change the vector potential only on a finite time scale (unlike the experimentally impractical $E$ field pulse discussed above). 
If now a pulse in $A(t)=A_0\delta(t-t_p)$ instead of $E(t)$ is applied (as is the case in the short pulse $a\to 0$ limit of the type-II pulse), the supercurrent is canceled almost instantly by a current with same magnitude but opposite direction after the vector potential is abruptly switched off again. The asymptotic long time behavior of the  system is not probed and (as in equilibrium) one can show that the imaginary part of the conductivity $\sigma(\omega)=j(\omega)/E(\omega)\sim j(\omega)/\omega A(\omega)$ shows a $1/\omega$ divergence with prefactor $\sim \Delta(t_p)$ in the adiabatic limit. The Fourier transform of the current reveals a behavior $\sim A_0 \Delta(t_p)$ at small $\omega$. Alternatively, one can use the integrated current $\int dt j(t)$ to probe the low frequency behavior as described above, which is equivalent to analyzing the low frequency divergence of the imaginary part of the conductivity. Of course this implies only that a change in the gap function would induce a change in the behavior of the imaginary part of the conductivity at small frequencies, but not necessarily the reverse.

\begin{figure}[t]
\centering
\includegraphics[width=\columnwidth]{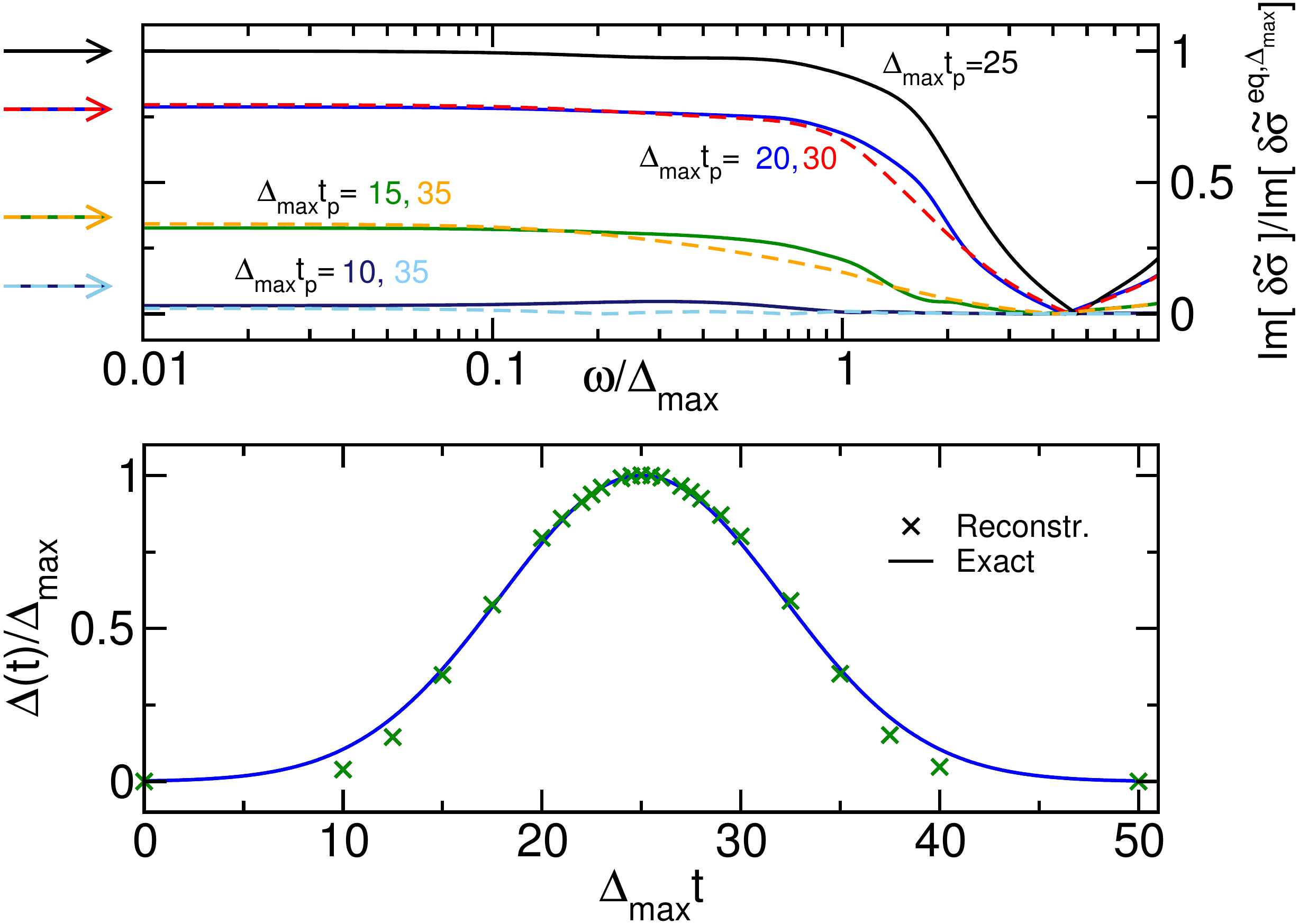}
\caption{Upper panel: the imaginary part of the difference in optical conductivity between the superconducting and the normal conducting state $\delta \tilde \sigma(\omega)=\tilde\sigma^{\rm s}(\omega)-\tilde\sigma^{\rm n}(\omega)$, relative to the same quantity in equilibrium evaluated at the maximum value of the gap. We choose $\Delta(t)=\Delta_2(t)$ and the = parameters as  $\Delta_{\rm ini}/\Delta_{\rm max}=0$, $\Delta_{\rm max}t_{0}=25$ and $\Delta_{\rm max}T_0=10$ and use a narrow type-II probe pulse $a\to 0$. Arrows to the left indicate the value of the instantaneous gap. Lower panel: Reconstructed values of the instantaneous gap compared to the exact $\Delta(t)$. The values are read off from the plateaus reached at low frequencies in $\delta \tilde \sigma(\omega)$.  }
\label{fig:A1shortim}
\end{figure}

We start with analyzing the probe of type-II in the short pulse limit. If the gap evolves sufficiently slow the system evolves perfectly adiabatically and simply the equilibrium optical conductivity is probed at time $t_p$.
The upper panel of Fig~\ref{fig:A1shortim} shows the imaginary part of the difference in optical conductivity between the superconducting and the normal conducting state $\delta \tilde \sigma(\omega)=\tilde\sigma^{\rm s}(\omega)-\tilde\sigma^{\rm n}(\omega)$, relative to the same quantity in equilibrium evaluated at the maximum value of the gap. We choose $\Delta(t)=\Delta_2(t)$ and the the parameters as  $\Delta_{\rm ini}/\Delta_{\rm max}=0$,  $\Delta_{\rm max}t_{0}=25$ and  $\Delta_{\rm max}T_0=10$. The low frequency behavior of the imaginary part of the optical conductivity reflects the behavior of the instantaneous gap for the slow gap variation  very well (exact values are given as arrows to the left of the plot). Analyzing the plateau value of the low frequency behavior of the imaginary part of the conductivity one can reconstruct the instantaneous gap as shown in the lower panel of Fig.~\ref{fig:A1shortim}.  We note that accessing very low frequencies is experimentally difficult. We see that the gap estimate requires access to frequencies of the order of 20\%  of the gap value.

\begin{figure}[t!]
\centering
\includegraphics[width=\columnwidth]{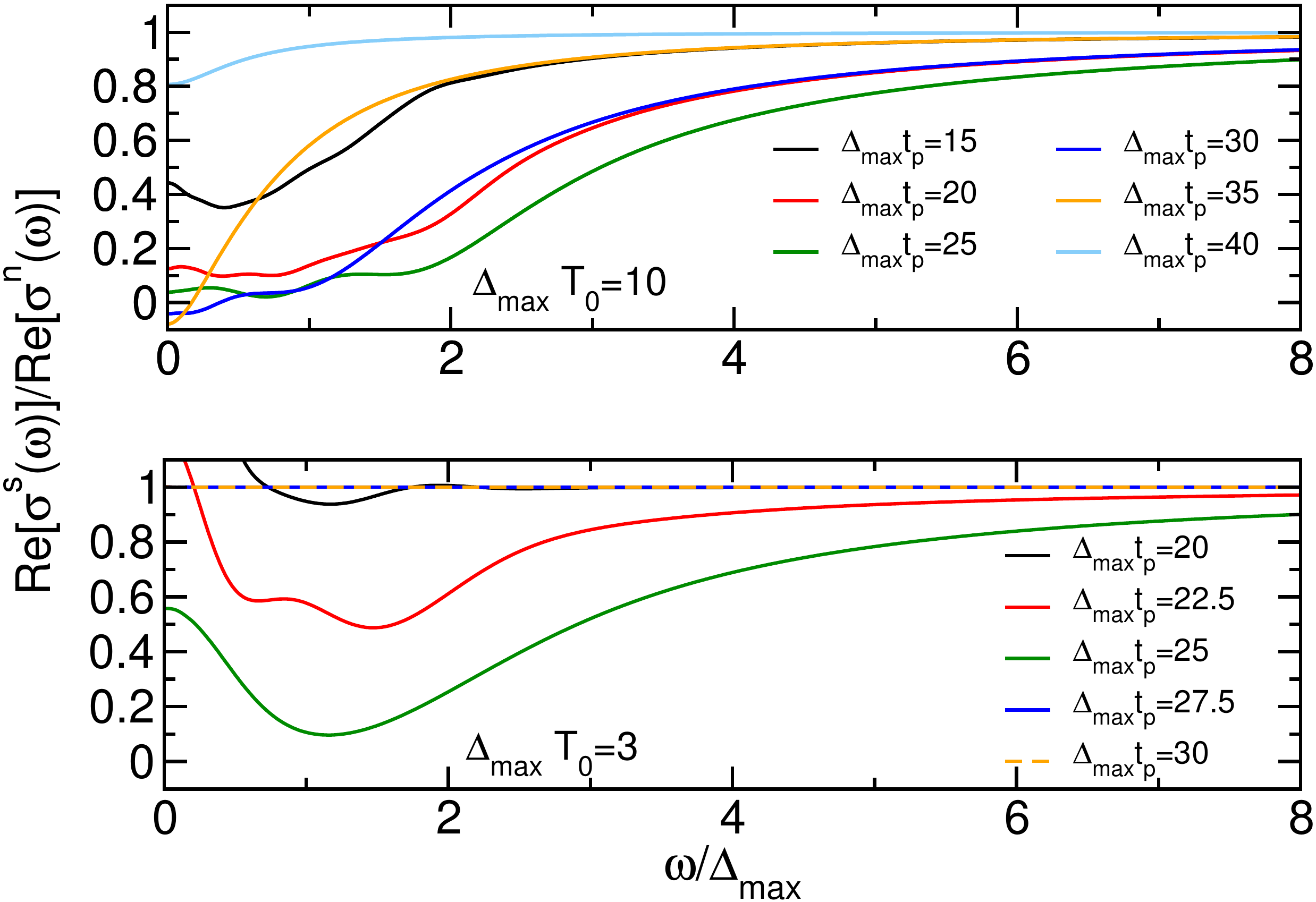}
\caption{Real part of the non-equilibrium optical conductivity. The parameters are the same as Fig.~\ref{fig:A1shortim} for both of the two values of $\Delta_{\rm max}T_0=10$ (upper panel) and $\Delta_{\rm max}T_0=3$ (lower panel).}
\label{fig:A1shortre}
\end{figure}

Next we analyze the real part of the conductivity. We use the same protocol for $\Delta(t)$ as well as for the probe as above. The results are summarized in Fig.~\ref{fig:A1shortre}. Compared to the equilibrium optical conductivity, several distinguishing features show up. First there is a pronounced oscillatory in gap feature, because the energy injected by the ramp is not instantly dissipated to the environment. This effect is more pronounced the less adiabatic the ramp is. Furthermore, there is an asymmetry between the measurements performed at $t_0+|x|$ and $t_0-|x|$ at low frequencies. If the probe is applied before the gap ramp $t_p=t_0-|x|$ the low frequency behavior of the conductivity probes the  future behavior of the gap. Thus the low-frequency behavior shows a hint of a suppression in the real part of the conductivity, if build-up of the gap is still to come, while it looks almost perfectly like the equilibrium normal conducting state if the ramp is in the past of the probe. This effect is the more pronounced the more non-adiabatic the ramp is. The in-gap contributions, attributed to the non-equilibrium nature of the system, makes a determination of the instantaneous gap $\Delta(t)$ form the real part of the conductivity difficult.       
   
We end our discussion of short probe pulses by shifting our interest to the type-I probe pulse, which in this limit implies a vector potential proportional to the derivative of a delta-function $A(t)\sim\delta'(t-t_p)$. A similar equilibrium analysis as done above for the type-II probe reveals that in the adiabatic limit the type-II probe can be used to determine  $p\Delta'(t_p)$ via the imaginary part of the conductivity. 

\subsection{Intermediate Width Time Probe-Pulses}

Finally we show the finite frequency part of the conductivity resulting from a type-I pulses (and also compare to the conductivity which would be found by the proposed type-II pulses). 

\begin{figure}[t]
\centering
\includegraphics[width=\columnwidth]{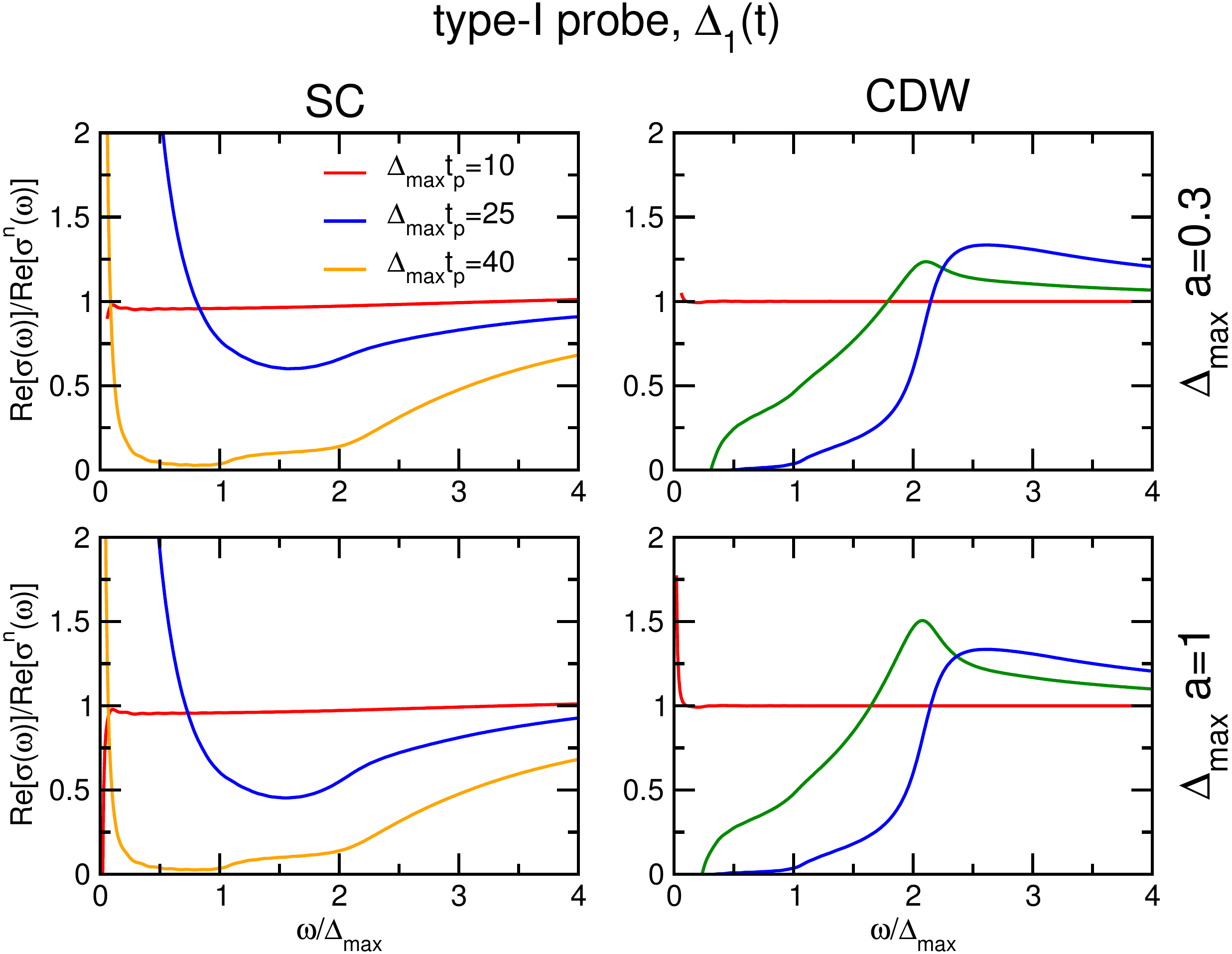}
\caption{Optical conductivity $\sigma(\omega)=j(\omega)/E(\omega)$ for a type-I probe pulse for $\Delta(t)=\Delta_1(t)$ with $\Delta_{\rm ini}/\Delta_{\rm max}=0$, $\Delta_{\rm max}t_0=25$, $\Delta_{\rm max}T_{0,{\rm rise}}=3$ and $\Delta_{\rm max}T_{0,{\rm fall}}\to \infty$ (corresponding to the left panels of Fig.~\ref{fig:Reconstruct}. We show the superconducting state (left panels) as well as a charge density wave state (right panels).  }
\label{fig:finwidthI}
\end{figure}

\begin{figure}[t]
\centering
\includegraphics[width=\columnwidth]{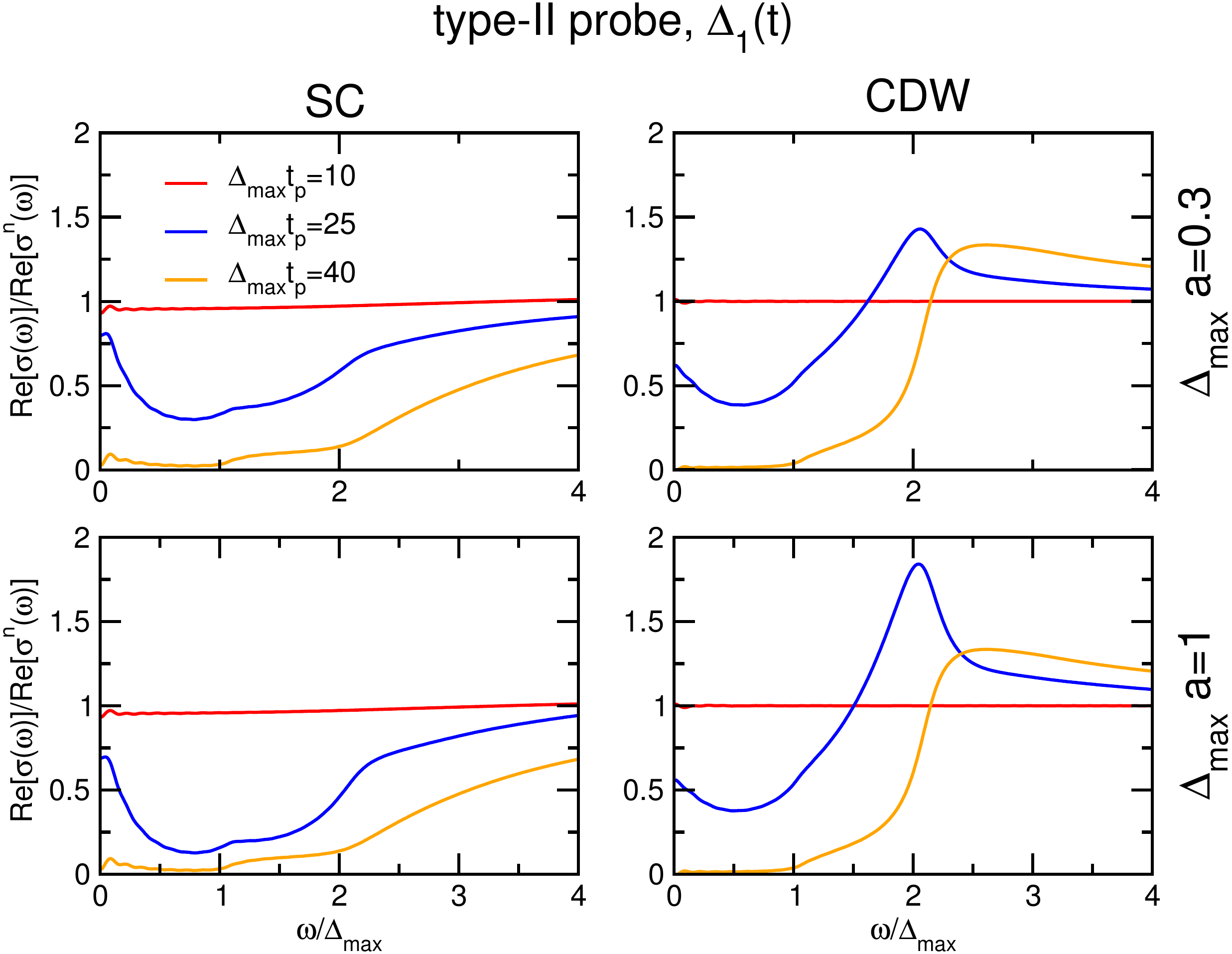}
\caption{Same as Fig.~\ref{fig:finwidthI} but for a type-II probe pulse. }
\label{fig:finwidthII}
\end{figure}

In Figs.~\ref{fig:finwidthI} and \ref{fig:finwidthII} we compare the measured optical conductivities $\sigma(\omega)=j(\omega)/E(\omega)$ for experimentally relevant type-I and proposed type-II probe pulses in the superconducting as well as the charge density wave state. We concentrate on a gap profile described by the left panels of Fig.~\ref{fig:Reconstruct}, so $\Delta(t)=\Delta_1(t)$ with  $\Delta_{\rm ini}/\Delta_{\rm max}=0$, $\Delta_{\rm max}t_0=0$, $\Delta_{\rm max}T_{0,{\rm rise}}=3$ and $\Delta_{\rm max}T_{0,{\rm fall}}\to \infty$. Type-II pulses seem to reproduce the equilibrium intuition slightly better as the small frequency behavior does not show the strong rise visible in the type-I case. The superconducting state seems to gradually evolve from the metallic initial state to the superconducting one. For the charge density wave state there is a pronounced enhancement of the edge at $\omega=2\Delta$ as the gap builds up.  

\begin{figure}[t]
\centering
\includegraphics[width=\columnwidth]{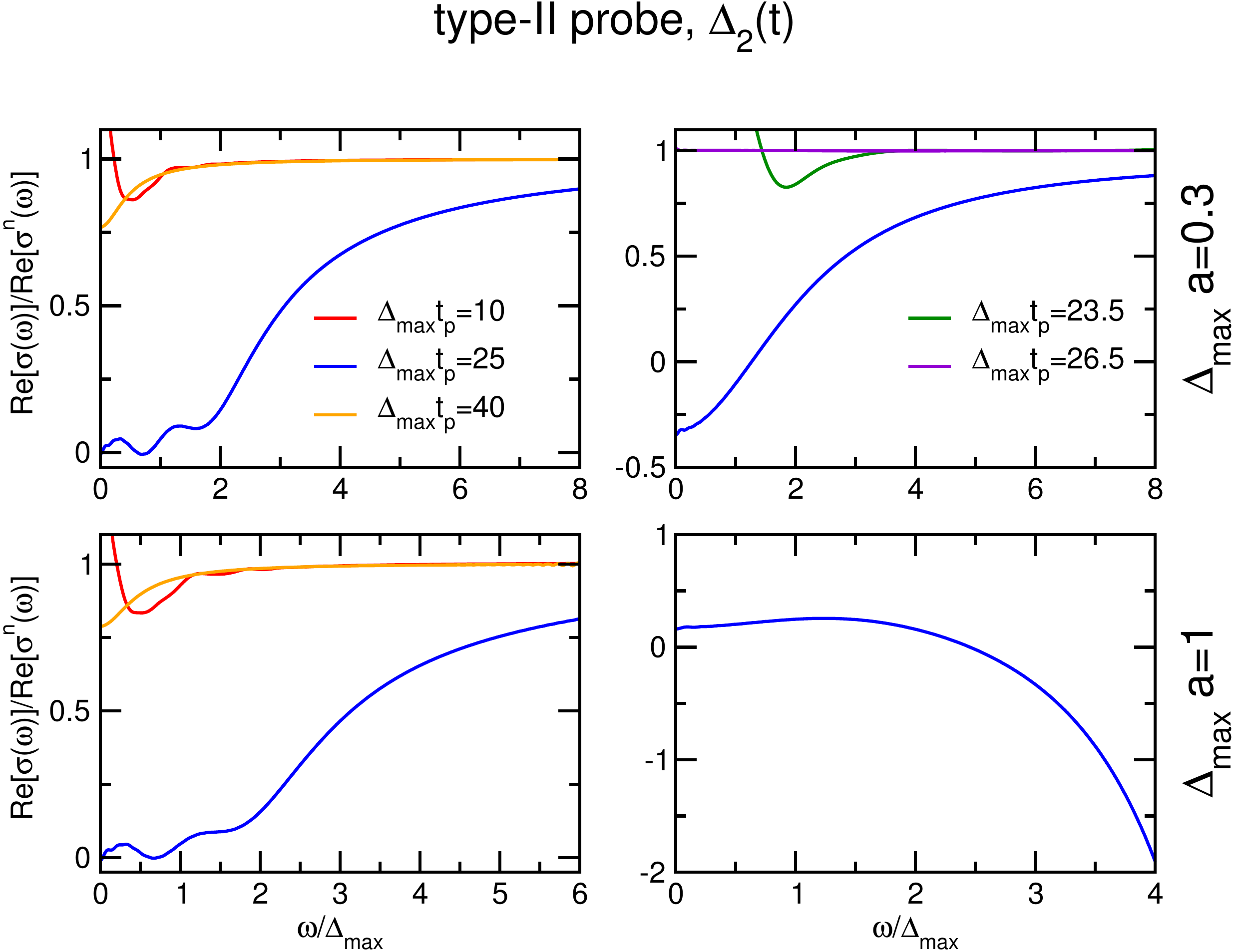}
\caption{ Same as Fig.~\ref{fig:finwidthI} but for $\Delta(t)=\Delta_2(t)$ with  $\Delta_{\rm ini}/\Delta_{\rm max}=0$, $\Delta_{\rm max}t_0=25$ and $\Delta_{\rm max}T_{0}=10$ (left panels) and $\Delta_{\rm max}T_{0}=1$ (right panels) and showing only results for the superconducting state.}
\label{fig:finwidthIII}
\end{figure}

Finally, in Fig.~\ref{fig:finwidthIII} we show how the optical conductivity of a short lived transient superconductor cannot be meaningfully reproduced by the same means. We concentrate on the time dependent gap profile shown in the middle panels of Fig.~\ref{fig:Reconstruct} and compare a slowly changing gap ($\Delta_{\rm max}T_0=10$) to a rapidly evolving profile ($\Delta_{\rm max}T_0=1$). In the slowly evolving case the equilibrium intuition works very well, with a gradual build up and vanishing of the optical gap in the conductivity. However, in the rapidly evolving case the reconstruction can deviate significantly from this intuition (the conductivity becoming negative or not resembling in anyway the equilibrium result).  

The reason for these deviations from equilibrium is two fold. There is a trivial contribution from the finite width of the probe pulse which measures the varying gap over a large time range. This contribution would yield differences to the equilibrium  optical conductivity even in the adiabatic limit and can be corrected, also  experimentally, by introducing delay lines between pump and probe pulse as documented in Refs. \onlinecite{Larsen11,Kindt99,Averitt00}. We are more interested in the second contribution which arises even for very short probe pulses and is linked to the intrinsic non-equilibrium state studied. We checked, that the negative feature in the conductivity arising even in the short pulse limit (upper right panel of Fig.~\ref{fig:finwidthIII}) can (at least partially) be traced back to the fact that the Fourier transform with respect to a single variable of a two-time function is not well defined. The additional lag of the current with respect to the probe field leads to a phase difference, which mixes real and imaginary part of the Fourier transforms $j(\omega)$ and $E(\omega)$ defining the optical conductivity. This admixture of imaginary part contributions to the real part leads to the conductivity being negative even in the short probe pulse limit.

\section{Conclusion}
 \label{summary} 
To summarize, we have presented a theory of the response of non-equilibrium  superconducting and charge density wave systems to weak applied ``probe'' electric fields.  The non-equilibrium state was introduced phenomenologically--we simply assumed a time varying order parameter $\Delta(t)$.  Our treatment was also based on a time-dependent mean field (BCS) approximation, supplemented by relaxation via a weakly coupled  fermionic reservior.  Note that in this approximation the gap function and the order parameter are identical and for the most part we have used the terms interchangeably, except that in Figures ~\ref{samplecond} and ~\ref{simulateexptA2_2} we use a phenomenological ansatz to investigate some features of the case in which phase fluctuations reduce the superfluid stiffness but not the gap. These simplifications were made to enable a focus on the issue of main interest, namely the observable features of the response to an applied probe field.  The extension of our work to  more general theoretical treatments  of a nonequilibrium state is straightforward and would be of interest, including in particular the inelastic (self-energy) effects arising in the nonequilibrium situation. 

Our work was motivated by the extensive and growing literature on optical studies of  transient behavior of electronic orders, and in particular by the discussion of Orenstein and Dodge \cite{Orenstein16} of the pitfalls inherent in   the common experimental practice of collapsing the theoretically expected two-frequency conductivity to a function of a single frequency. We investigate time-domain signatures of the important physics, in particular  identifying a characteristic difference in the time domain responses of transient  superconducting and  charge density wave states: the initial current response in the two cases is opposite with respect to normal state current response.  We also  show how appropriate tailoring of pulses can enhance the signatures of the key physics of transient states, in particular revealing the instantaneous superfluid stiffness.

This work opens the route towards many intriguing further investigations. One obvious extension is to describe the dynamical melting of charge density wave order in recent Terahertz light field experiments.\cite{cdw1} and dynamical destruction of Mott insulators.  On the theoretical side, relaxing the time-dependent BCS assumption made here is an important direction. In particular we have assumed that at all relevant times the phase of the superconducting order parameter is spatially uniform and that the phase stiffness is large enough that the standard linear response to vector potential arguments apply. Relaxing these assumptions by incorporating random initial phases and the associated gradual annealing of the phase difference over time after the superconducting quench is of interest, as is considering the potential for nonlinear response at early times when the superconducting order parameter is small. 
 
\section*{Acknowledgement}
We thank Dmitri Basov, Richard Averitt and Joseph Orenstein for valuable discussions.  
DMK was supported by DFG KE 2115/1-1. AJM and EYW were supported by the Basic Energy Sciences Program of the US Department of Energy under grant \text{DE-SC}0012592.  DRR was supported by NSF CHE-1464802.  \\

\appendix*
\setcounter{equation}{0}
\section*{Appendix}
\label{appen} 
 The non-equilibrium Green's functions $G^{\rm R/K/A}(k,t,t')$ are calculated for a given gap $\Delta(t)$ by solving 
\begin{align}
i\partial_t G^{\rm R}(k,t,t')&=H^{\rm 1P}(t)G^{\rm R}(k,t,t')\;\;\;\;\;t>t'\label{eq:Gr_det}\\
G^{\rm A}(k,t,t')&=[G^{\rm R}(k,t',t)]^*\\
G^{\rm K}(k,t,t')&=-iG^{\rm R}(k,t,t_{\rm ini})(1-2n_{0,k})G^{\rm A}(k,t_{\rm ini},t')
\end{align}
where $H^{\rm 1P}$ is the time dependent one-particle Hamiltonian given by the two by two matrix in Eq.~\eqref{eq:Hsc}. The initial conditions are given by $G^{\rm R}(k,t,t)=-i$ as well as by the densities $n_{0,k}=\left\langle\Psi^\dagger_k \Psi_k\right\rangle$ at the initial time $t_{\rm ini}$. We choose as the initial state from now on the ground state of the system. This determines the occupancies $n_{0,k}$. We solve Eq.~\eqref{eq:Gr_det} by discretizing time into very small steps $\Delta t$ and iteratively propagating $G^{\rm R}(k,t+\Delta t,t')=e^{-i\bar H \Delta t}G^{\rm R}(k,t,t')$ by using the  piece-wise constant mid-point Hamiltonian $\bar H=[H(t+\Delta t)+H(t)]/2$. We choose a flat density of states (linear dispersion $\epsilon_k$) of width $D=20\Delta_{\rm max}\gg \Delta_{\rm max}$ in our calculations. 

Before using these Green's functions in the calculation of the optical conductivity we have to consider a final complication: the non-equilibrium ramp of $\Delta(t)$ will inject energy into the system. This energy will never dissipate if the dynamics is given by Eq.~\eqref{eq:Hsc}. Furthermore, the system is non-interacting and thus we do not include any mechanism which could relax energy among the modes described by $k,\sigma$. Thus even at asymptotic long times the system will never thermalize. A more realistic model of a condensed matter system would include a coupling between the system and its environment, which can act as a thermal bath. The bath introduced for this purpose should not introduce a significant source of scattering changing the properties of the system. Here we couple the system in k-space to independent one-dimensional tight-binding reservoirs with bandwidth $D_{\rm res}$ of length $L$, but the precise choice of bath is not relevant to our conclusions.  This modifies the single particle Hamiltonian to be used in  Eq.~\eqref{eq:Gr_det} from the two by two matrix $H^{\rm 1P}$ to the $2L+2$ by $2L+2$ matrix
\begin{equation}
H^{\prime\rm 1P}=
\begin{pmatrix}
H^{\rm res, 1P}&K_c^T&0&0\\
K_c&H^{\rm 1P}_{1,1}&H^{\rm 1P}_{1,2}&0\\
0&H^{\rm 1P}_{2,1}&H^{\rm 1P}_{2,2}&M_c\\
0&0&M_c^T&H^{\rm res, 1P}
\end{pmatrix}
\end{equation}
where $H^{\rm res, 1P}$ is the $L$ by $L$ matrix
\begin{equation}
H^{\rm res, 1P}=
\begin{pmatrix}
0&D_{\rm res}/4&&\\
D_{\rm res}/4&\ddots&\ddots&\\
&\ddots&\ddots&D_{\rm res}/4\\
&&D_{\rm res}/4&0\\
\end{pmatrix}
\end{equation}
and
\begin{align}
K_c&=\begin{pmatrix}
0&\dots&0&t_c
\end{pmatrix}\\
M_c&=\begin{pmatrix}
t_c&0&\dots&0
\end{pmatrix}
\end{align}
We choose $L$ large enough, such that the finiteness of the additional reservoirs can be neglected. This procedure introduces a decay mechanism to the Green's functions without spoiling their physical properties. We define the physically relevant hybridization $\Gamma=4t_c^2/D_{\rm res}$ which characterizes the decay rate introduced by the additional reservoir. We choose $D_{\rm res}=4\Delta_{\rm max}$ and $\Gamma=1/9\Delta_{\rm max}$ in the following whenever not explicitly written otherwise.
\begin{figure}[t]
\centering
\includegraphics[width=\columnwidth]{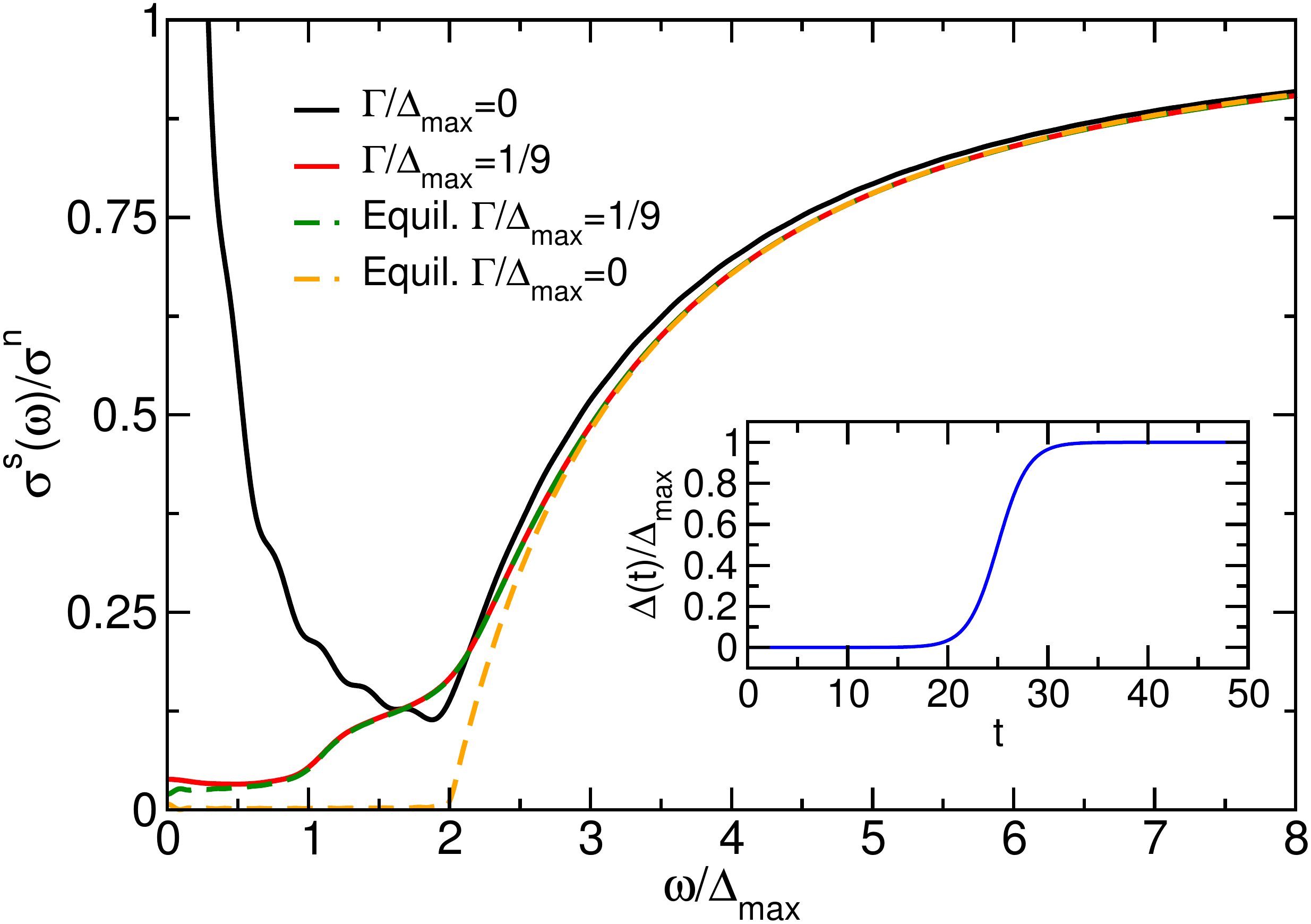}
\caption{Steady-state optical conductivity as measured by a very late electric field pulse $E(t)=E_0\delta(t-t_p)$. The input gap is of the form $\Delta_1(t)$ with $\Delta_{\rm max}T_{0,{\rm fall}}\to \infty$ and we choose the other parameters as $\Delta_{\rm ini}/\Delta_{\rm max}=0$, $\Delta_{\rm max}t_0=25$, $\Delta_{\rm max}T_{0,{\rm rise}}=3$ and $\Delta_{\rm max}t_p=50$. For comparison we show the equilibrium result as well. Inset: Gap profile used to perform the calculations.}
\label{fig:thermasig}
\end{figure}

To illustrate the necessity of reservoirs in the description of the optical conductivity, we consider a superconductor with time dependent gap of the form $\Delta_1(t)$ with $\Delta_{\rm max}T_{0,{\rm fall}}\to \infty$ and we choose the other parameters as $\Delta_{\rm ini}/\Delta_{\rm max}=0$,  $\Delta_{\rm max}t_0=25$ and $\Delta_{\rm max}T_{0,{\rm rise}}=3$ . The gap continuously rises from zero to one (compare inset of Fig.~\ref{fig:thermasig}). If we measure the optical conductivity at large times the system has relaxed to a steady state and the two times optical conductivity depends only on its relative time argument $\sigma(t,t')=\sigma(t-t')$. Taking the Fourier transform with respect to this relative time argument connects to the well understood equilibrium results. Figure~\ref{fig:thermasig} shows the steady-state optical conductivity $\sigma(\omega)$ in Fourier space with ($\Gamma/\Delta_{\rm max}=1/9$) and without ($\Gamma/\Delta_{\rm max}=0$) coupling the BCS superconductor to an environment. We measure the optical conductivity by an electric field pulse $E(t)=E_0\delta(t-t_p)$ applied at time $\Delta_{\rm max}t_p=50$ well after the gap reaches its plateau. The value of $\Delta_{\rm max}t_p=50$ is large enough such that the optical conductivity has become approximately steady. The decoupled case clearly shows a deviation from the equilibrium prediction. There is a pronounced oscillatory in-gap content in the non-equilibrium steady state, which is absent in the equilibrium case. The excess energy injected by the ramp in the gap $\Delta(t)$ is neither dissipated nor redistributed in a thermal fashion.  However, with environment coupling the steady state prediction agrees with the equilibrium prediction of the same system. The reservoir thus provides a meaningful thermalization mechanism.

{}


\begin{thebibliography}{}
\bibitem{footnote1}
In the non-equilibrium theory we propose we assume a given time varying gap $\Delta(t)$. The phase of the order parameter $\Delta(t)$ is thus homogeneous in space. Studying the consequences of this assumptions of perfect phase stiffness is  subject of future work.
\end{thebibliography}
\end{document}